\documentclass[pdflatex,sn-mathphys-num]{sn-jnl}
\usepackage{geometry}

\usepackage{graphicx}
\usepackage{amsmath,amssymb}
\usepackage{booktabs}
\usepackage{subcaption}
\usepackage{array}
\usepackage{tabularx}
\usepackage{caption}
\begin{document}

\title
{Structural and Optical Properties of PVP-Modified ZnS Nanoparticles for Photocatalytic Dye Degradation}

\author[1]{\fnm{Vigneshwari} \sur{S R}}

\author[1]{\fnm{Dushyant} \sur{Singh}}

\author[1]{\fnm{Subha} \sur{Maity}}

\author[1]{\fnm{Tharundev} \sur{V V}}

\author[2]{\fnm{Mukesh K.} \sur{Kumawat}}

\author[2]{\fnm{Mukesh} \sur{Dhanka}}

\author*[1]{\fnm{Krista R.} \sur{Khiangte}} \email{Krista.khiangte@iitgn.ac.in}

\affil[1]{\orgdiv{Department of Physics},
  \orgname{Indian Institute of Technology Gandhinagar},
  \orgaddress{\city{Gandhinagar}, \postcode{382355},
  \state{Gujarat}, \country{India}}}
  
  \affil[2]{\orgdiv{Department of Biological Sciences and Engineering},
  \orgname{Indian Institute of Technology Gandhinagar},
  \orgaddress{\city{Gandhinagar}, \postcode{382355},
  \state{Gujarat}, \country{India}}}

\abstract{Pristine and PVP-modified ZnS nanoparticles were investigated to examine how PVP addition affects structural, surface, optical, and methylene blue (MB) decolorization properties. X-ray diffraction confirmed cubic zinc-blende ZnS in all samples, with apparent Scherrer coherent-domain sizes of 3.10--3.31 nm. Electron microscopy revealed agglomeration, while HRTEM and electron diffraction confirmed locally crystalline regions in S1 and S4. The estimated direct Tauc gap decreased from 3.69 eV for S1 to 3.58 eV for S4. Between 100 and 300 K, the principal TO-like Raman band softened on heating, while visible photoluminescence changed predominantly within the overlapping green--yellow region. After 210 min under 254 nm irradiation, MB-band losses near 664 nm reached 19.33\%, 58.18\%, 70.00\%, and 94.42\% for S1--S4, compared with 6.58\% for the catalyst-free control. PVP modification therefore alters optical and dye-decolorization responses while preserving the principal ZnS phase; however, dye-band attenuation alone does not establish complete mineralization.}


\maketitle

\section{Introduction}
\label{sec:introduction}

Zinc sulfide (ZnS) is a direct-wide-band-gap II--VI semiconductor whose
electronic, optical and vibrational properties are sensitive to crystal
structure, particle size and surface chemistry. Its cubic zinc-blende
and hexagonal wurtzite forms provide different local structural
environments, while nanoscale dimensions introduce additional effects
associated with confinement and surface coordination. These properties
make ZnS relevant to ultraviolet photodetection, luminescent materials
and photochemical applications. Fang \textit{et al}.\ demonstrated
visible-light-blind ultraviolet detection using single-crystalline ZnS
nanobelts \cite{Fang2009ZnSUVSensors}, while studies of ZnS-based
electroluminescent phosphors have established the importance of the
local environments of optically active centres
\cite{Warkentin2007ZnSElectroluminescence}. ZnS therefore provides a useful system for examining how changes in structure and surface bonding influence its electronic structure and optical properties. Changes in composition and preparation conditions can further modify these properties in nanocrystalline ZnS, as illustrated by studies of Fe- and Co-modified samples \cite{VigneshwariFeZnS2026,VigneshwariCoZnS2026}.

As particle dimensions decrease, quantum confinement can modify the electronic transitions, while the increasing surface-to-volume ratio strengthens the influence of under-coordinated atoms, defects and interfacial disorder. These contributions can coexist, so a change in the absorption edge or PL spectrum cannot necessarily be attributed to particle size alone. Wang \textit{et al}.\ demonstrated that different defect environments in ZnS nanocrystals produce distinct visible-emission features, illustrating the sensitivity of luminescence to preparation and surface treatment \cite{Wang2011ZnSDefectEmission}. The same localized states can influence carrier trapping and recombination, which also affect photocatalytic activity. Photocatalytic pollutant transformation has been demonstrated using ZnS nanocrystals, and subsequent studies have examined how defect engineering modifies their optical response, activity and photostability \cite{TorresMartinez2001ZnSPhotocatalysis,Hao2018ZnSVacancyPhotostability,Jubeer2023Defect}. Understanding these properties requires consideration of both the crystalline regions and their surrounding interfaces.

Organic capping agents offer a practical means of controlling nanoparticle formation and modifying the semiconductor surface. Polymers such as polyvinyl alcohol (PVA), polyethylene glycol (PEG) and polyvinylpyrrolidone (PVP), surfactants such as cetyltrimethylammonium bromide (CTAB), and thiol-containing ligands have been used in ZnS preparation. Their different functional groups and interactions with precursor species can alter nucleation, growth, aggregation and surface coordination. Deka \textit{et al}.\ compared PVA-, PVP- and CTAB-capped ZnS nanocrystals and reported changes in structural, optical and photocatalytic properties with the choice of capping agent \cite{Deka2019ZnSCapping}. Tiwari \textit{et al}.\ found different PL responses for ZnS nanoparticles capped with L-cysteine, mercaptoethanol and mercaptopropionic acid \cite{Tiwari2015ZnSThiolCapping}. These studies establish that capping agents influence the physical properties of ZnS through both growth-related effects and changes in surface chemistry.

PVP is particularly useful because its carbonyl-containing pyrrolidone groups can interact with Zn-containing species and accessible surface Zn sites, while its polymer chains provide steric stabilization. Ghosh \textit{et al}.\ investigated the synthesis and optical characterization of PVP-encapsulated ZnS nanoparticles, and Soltani \textit{et al}.\ demonstrated that PVP concentration influences the particle size and dispersion of microwave-prepared ZnS \cite{Ghosh2006PVPZnS,Soltani2012PVPZnS}. Kumar \textit{et al}.\ also compared PEG- and PVP-capped ZnS in photocatalytic pollutant removal, showing that polymer modification can affect the response under irradiation \cite{Kumar2022PVPCapped}. These findings motivate the use of PVP as a controllable modifier, although they do not establish its universal superiority over other capping agents. Its effects depend on the preparation conditions and amount added. Surface stabilization and passivation may modify recombination pathways, whereas greater polymer coverage may restrict access to reactive sites or alter interfacial charge transfer. Varying the nominal PVP content therefore provides a means of examining the balance between changes in nanoparticle formation, optical response and surface reactivity.

Temperature-dependent Raman and photoluminescence (PL) spectroscopy
provide complementary probes of this evolution. Raman peak positions
and linewidths respond to thermal expansion, phonon anharmonicity,
disorder and strain. Serrano \textit{et al}.\ investigated the Raman spectra
of cubic ZnS and the mechanisms influencing its optical-phonon
linewidths \cite{Serrano2004ZnSRaman}. Surface-phonon studies of
wurtzite ZnS nanowires \cite{Xiong2004ZnSSurfacePhonons} and
surface-sensitive vibrational measurements on nanoscale cubic ZnS
\cite{Juine2020} further demonstrate the role of
geometry and the interface in the vibrational response. The
anharmonic framework developed by Balkanski \textit{et al}.\ for silicon
provides a general basis for interpreting temperature-dependent
phonon shifts and broadening
\cite{Balkanski1983AnharmonicPhonons}. PL probes electronic relaxation and emission
redistribution. Comparing PL and Raman across the ZnS series provides
complementary insight into electronic emission and lattice dynamics.

The preceding studies motivate a systematic investigation of how
controlled PVP addition influences the structure, absorption edge
and temperature-dependent Raman and PL responses of ZnS, together
with its dye-decolorization behaviour. Changes in particle size or
room-temperature PL alone do not establish the thermal evolution
of these properties, while enhanced decolorization does not identify
the underlying carrier-relaxation pathways. We hypothesize that
PVP-assisted preparation modifies nanoparticle growth and surface
coordination, producing distinct optical and vibrational responses
while retaining the principal ZnS phase. Comparing these responses
within a common sample series allows their dependence on PVP content
to be examined alongside changes in UV-assisted dye decolorization.

In this work, pristine ZnS and PVP-modified ZnS nanoparticles prepared with PVP additions of 0, 0.5, 1.5 and 2.5 g for S1–S4 respectively are investigated. X-ray diffraction, FTIR spectroscopy, electron microscopy and elemental analysis are used to examine their crystalline structure, surface-related vibrational features, morphology and composition. UV--visible spectroscopy is employed to evaluate changes in the absorption edge and estimated optical gap, while Raman and PL measurements between 100 and 300~K probe the thermal evolution of lattice vibrations and broad visible emission. The UV-assisted decolorization of methylene blue (MB) is evaluated under 254~nm irradiation, including dark-contact measurements and a catalyst-free irradiation control. The analysis distinguishes attenuation of the characteristic MB absorption band from complete mineralization, since dye transformation can alter both the intensity and profile of the visible absorption spectrum \cite{Zhang2001MBDemethylation}. Together, these measurements examine how PVP-assisted preparation affects the structural, optical and vibrational properties of ZnS and how these changes accompany its dye-decolorization response.

\section{Experimental Details}

This section describes the experimental procedures employed for the synthesis
and characterization of pristine and PVP-capped ZnS nanoparticles (NPs). The
methodology is organized into three parts: the materials and reagents used, the
synthesis routes adopted for pristine ZnS and samples containing different PVP
concentrations, and the characterization techniques employed to examine their
structural, vibrational, chemical, and optical properties. The precursor
processing and post-synthesis treatment were adapted from previously reported
ZnS preparation protocols, with transition-metal incorporation omitted and PVP
introduced as a controlled surface modifier
\cite{VigneshwariFeZnS2026,VigneshwariCoZnS2026}.

\subsection{Materials}

Zinc chloride (ZnCl$_2$, 97\% extra pure, dry),
sodium sulfide hydrate flakes (60\% extra pure),
and polyvinylpyrrolidone (PVP, K30 grade) were
purchased from Loba Chemie Pvt. Ltd. (Mumbai,
India). ZnCl$_2$ and sodium sulfide served as the
zinc and sulfur precursors, respectively, while
PVP was used as a surface-capping agent during
the synthesis of ZnS nanoparticles. Deionized
(DI) water was used as the reaction medium and
for preparing the precursor solutions. Methylene
blue (MB) was used as the model organic dye for
the photocatalytic experiments. All chemicals
were used as received without further
purification.

\subsection{Preparation of Pristine and PVP-Modified ZnS Nanoparticles}

Pristine and PVP-modified ZnS nanoparticles were prepared by aqueous
precipitation, with precursor preparation and post-synthesis treatment
adapted from previously reported procedures
\cite{VigneshwariFeZnS2026,VigneshwariCoZnS2026}. For each synthesis,
20~mL of a 0.10~M ZnCl$_2$ solution was prepared in DI water and
stirred at 45~$^\circ$C for 20~min. Separately, 10~mL of a 0.2~M
Na$_2$S solution was prepared in deionized water and stirred under the
same conditions.

For pristine ZnS (S1), the Na$_2$S solution was added dropwise to the
ZnCl$_2$ solution under continuous stirring. After addition, stirring
was continued to allow further reaction. A white precipitate formed,
consistent with ZnS precipitation according to
\begin{equation}
\mathrm{ZnCl_2(aq) + Na_2S(aq)
\longrightarrow ZnS(s) + 2NaCl(aq)}.
\label{eq:zns_formation}
\end{equation}

For the PVP-modified samples, the precursor concentrations, solution
volumes and reaction temperature were kept unchanged. Before adding
Na$_2$S, 0.5, 1.5 and 2.5~g of PVP were introduced into the ZnCl$_2$
solution for S2, S3 and S4, respectively. Each polymer-containing
solution was stirred for 60~min at 45~$^\circ$C, after which the
Na$_2$S solution was added dropwise using the procedure described
for S1. Introducing PVP before precipitation allows the polymer
to interact with the precursor species and developing nanoparticle
surfaces, as reported in previous PVP-assisted ZnS syntheses
\cite{Ghosh2006PVPZnS,Soltani2012PVPZnS}.

The precipitates were collected by centrifugation and washed repeatedly
with deionized water and methanol to remove soluble salts, unreacted
precursors and excess polymer. Each product was then heat-treated at
200~$^\circ$C for 2~h, following the post-synthesis treatment used in
the related ZnS preparations
\cite{VigneshwariFeZnS2026,VigneshwariCoZnS2026}. After cooling, the
products were gently ground to obtain powders. The samples were
designated S1--S4 in order of increasing PVP addition.

\subsection{Preparation of Methylene Blue Solution}

Methylene blue (MB) was used as the model dye in the photocatalytic measurements. An aqueous MB solution with an initial concentration of 5~mg\,L$^{-1}$ (approximately 5~ppm) was prepared in DI water to a final volume of 100~mL.

\subsection{Characterization}

The structural, chemical, optical, and vibrational properties of S1--S4 were examined using complementary characterization techniques. High-resolution X-ray diffraction (HRXRD) measurements were performed using a Rigaku SmartLab diffractometer operated at an accelerating voltage of 30~kV, with Co K$\alpha$ radiation ($\lambda_{Co} = 1.7906$~\AA) as the X-ray source. The measurements were used to identify the crystalline phase and evaluate changes in the diffraction features of the synthesized ZnS nanoparticles. Fourier-transform infrared (FTIR) spectroscopy was used to identify the principal chemical bonds and surface functional groups present in the samples. Scanning electron microscopy (SEM) was employed to investigate the surface morphology and particle characteristics of the synthesized nanoparticles. The measurements were carried out using a JEOL JSM-7600F field-emission scanning electron microscope, equipped with an energy-dispersive X-ray spectroscopy (EDS) system for elemental analysis. The internal nanostructure and crystallinity of the samples were further investigated using high-resolution transmission electron microscopy (HRTEM) operated at an accelerating voltage of 300~kV. The measurements were performed using an FEI Themis 60-300 transmission electron microscope system.

UV--visible absorption spectra were recorded using a UV--Vis spectrophotometer of Agilent Technologies to determine the absorption-edge behaviour and optical band gap of each sample. UV--visible absorption spectroscopy was also used to monitor the methylene-blue absorption band near 664~nm during the photocatalytic measurements. Spectra were recorded from 500 to 800~nm after separating the photocatalyst particles from the withdrawn solution. Photoluminescence (PL) measurements were carried out using a Renishaw InVia Reflex Confocal Raman spectrometer equipped with a 325nm excitation laser to investigate the visible-emission characteristics and defect-related recombination processes. Temperature-dependent PL spectra were recorded from 300 to 100~K at intervals of 20~K to examine changes in emission intensity, peak position, and spectral profile.

Raman spectroscopy was carried out using a Renishaw inVia Reflex
confocal Raman spectrometer equipped with an 830~nm excitation
laser and a diffraction grating of 1200~lines\,mm$^{-1}$.
Temperature-dependent Raman spectra were recorded from 300 to
100~K in steps of 20~K to examine the vibrational modes and
lattice dynamics of ZnS. The results obtained from these techniques were subsequently analyzed in their respective sections to establish correlations among the crystalline structure, morphology, chemical environment, elemental composition, electronic transitions, defect emission, phonon behaviour, and photocatalytic performance of the synthesized nanoparticles.

\section{Results and Discussion}
\subsection{Fourier Transform Infrared Spectroscopy Analysis}
\label{subsec:ftir}

\begin{figure}[htbp]
    \centering
    \includegraphics[width=0.8\linewidth]{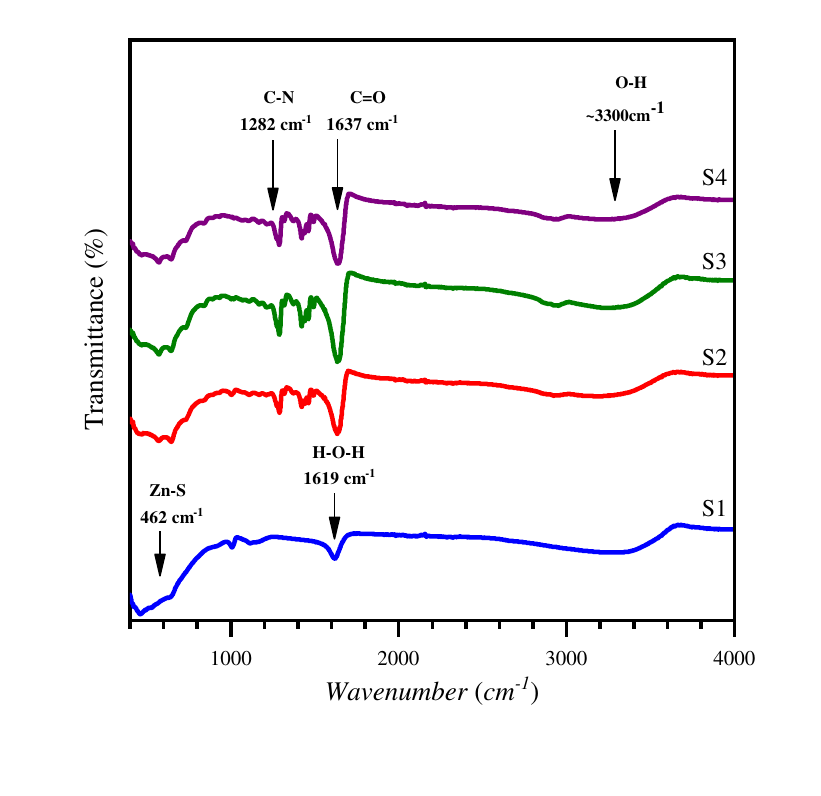}
    \caption{FTIR transmittance spectra of pristine ZnS (S1) and
    PVP-modified ZnS nanoparticles (S2--S4), recorded over
    400--4000~cm$^{-1}$.}
    \label{fig:ftir_spectra}
\end{figure}

Fourier transform infrared (FTIR) spectroscopy was used to identify
surface-bound molecular species in pristine ZnS (S1) and PVP-modified
ZnS (S2--S4). The analysis focuses primarily on the vibrational
fingerprint of PVP and on bands associated with adsorbed water. FTIR
provides complementary information on the chemical environment of the
samples, while the crystalline phase is determined more reliably by
X-ray diffraction \cite{Coates2000,Koczkur2015PVP}. The transmittance
spectra recorded between 400 and 4000~cm$^{-1}$ are shown in
Fig.~\ref{fig:ftir_spectra}. Successive spectra are vertically
stacked for clarity.

The S1 spectrum contains a broad absorption band extending through
approximately 3000--3600~cm$^{-1}$, with a minimum near
3317~cm$^{-1}$, together with a band near 1619~cm$^{-1}$. These
features are assigned to O--H stretching and H--O--H bending,
respectively, arising from molecular water adsorbed on the
nanoparticle powder. Similar features have been reported for ZnS
nanoparticles exposed to ambient moisture
\cite{jothibas2017structural}. Their presence does not imply that
oxygen is incorporated substitutionally into the ZnS lattice.

S1 also shows weak minima near 1119 and 1006~cm$^{-1}$. Bands in this
region can arise from residual precursor-derived or surface-adsorbed
species, but they are not sufficiently specific for a unique
assignment in the absence of additional chemical analysis. These
features are therefore retained as unassigned fingerprint-region
bands. In particular, the present spectrum does not show a clearly
resolved minimum near 921~cm$^{-1}$.

The low-wavenumber region differs between the samples. S1 exhibits a
broad minimum near 462~cm$^{-1}$, whereas S2--S4 show additional
structure near 570 and 641~cm$^{-1}$. Absorptions in this region have
been associated with ZnS-related lattice, surface, or multiphonon
vibrations in nanoparticle powders
\cite{jothibas2017structural}. 

The clearest difference between S1 and S2--S4 is the appearance of
bands characteristic of PVP. The PVP-modified samples exhibit
absorption bands near 1281--1283, 1427--1429, and
1635--1639~cm$^{-1}$, together with a weaker band near
2929--2948~cm$^{-1}$. The band near 1282~cm$^{-1}$ is assigned
primarily to C--N stretching of the PVP pyrrolidone ring. The feature
near 1428~cm$^{-1}$ is assigned to the coordination interaction
between the nitrogen atom of PVP and Zn$^{2+}$ ions, consistent with
the FTIR assignment reported in the literature
\cite{ankinapalli2024synthesis}. The weak band around
2940~cm$^{-1}$ is associated with aliphatic C--H stretching
\cite{Ghosh2006PVPZnS,Soltani2012PVPZnS,Shahi2017}. The simultaneous presence of
these PVP-related bands in S2--S4, together with their absence or
substantial suppression in S1, supports the retention of PVP in the
modified samples after synthesis and washing.

The intense band at 1635--1639~cm$^{-1}$ is assigned predominantly
to the C=O stretching vibration of the PVP pyrrolidone group.
Measurements that include a free-PVP reference place the carbonyl
band of uncoordinated PVP near 1654--1655~cm$^{-1}$ and report its
displacement to approximately 1633--1636~cm$^{-1}$ after the
formation of PVP-capped ZnS nanoparticles
\cite{Soltani2012PVPZnS,Shahi2017}. The position measured here is consistent
with these reported ZnS--PVP spectra. Interaction of the polar
pyrrolidone group with coordinatively unsaturated Zn sites can modify
the electronic environment of the carbonyl group and lower its
effective stretching frequency. Published studies further indicate
that both the C=O and C--N environments can participate in the
interaction between PVP and the ZnS surface
\cite{Ghosh2006PVPZnS,Soltani2012PVPZnS,Koczkur2015PVP}.

The evidence for PVP
association rests on the agreement of the observed
1635--1639~cm$^{-1}$ band with published ZnS--PVP spectra and,
more importantly, on its occurrence together with the C--N and C--H
bands. The carbonyl envelope may also contain a contribution from the
H--O--H bending vibration of adsorbed water, because this vibration
occurs near 1620--1640~cm$^{-1}$ and the broad O--H stretching band
remains visible in all four samples.

\begin{table}[htbp]
\centering
\caption{Principal FTIR bands obtained from the measured spectra and
their proposed assignments.}
\label{tab:ftir_assignments}
\small
\begin{tabular}{p{0.24\linewidth}p{0.20\linewidth}p{0.45\linewidth}}
\hline
Wavenumber (cm$^{-1}$) & Samples & \multicolumn{1}{c}{Assignment} \\
\hline

$\ 462$ &
S1 &
Low-frequency ZnS-related, surface, or multiphonon response;
assignment not unique \\

570, 641 &
S2--S4 &
Low-frequency envelope containing possible ZnS-related and
surface/PVP contributions; not used for phase identification \\

1004--1010 and 1106--1119 &
S1--S4 &
Weak fingerprint-region features; not uniquely assigned \\

1281--1283 &
S2--S4 &
Predominantly C--N stretching of the PVP pyrrolidone ring \\

1427--1429 & S2--S4 &
Coordination interaction between the nitrogen atom of PVP and Zn$^{2+}$ ions
\cite{ankinapalli2024synthesis} \\

$\ 1619$ &
S1 &
H--O--H bending of adsorbed molecular water \\

1635--1639 &
S2--S4 &
Predominantly C=O stretching of surface-associated PVP, with
possible overlap from adsorbed-water bending \\

2929--2948 &
S2--S4 &
Aliphatic C--H stretching of PVP \\

3180--3317 &
S1--S4 &
Broad hydrogen-bonded O--H stretching, mainly from adsorbed water \\

\hline
\end{tabular}
\end{table}

PVP can limit nanoparticle coalescence by adsorbing through its polar
pyrrolidone groups, while the polymer backbone extends away from the
surface and provides steric resistance between neighbouring particles
\cite{Ghosh2006PVPZnS,Koczkur2015PVP,Soltani2012PVPZnS}. The FTIR results establish
the presence of PVP in S2--S4 and are consistent with its
surface-capping role. However, FTIR alone does not determine the
surface coverage, binding stoichiometry, or absolute amount of
retained polymer.

The relative FTIR band depths should not be used to infer the amount
of PVP in S2--S4. 

Overall, the FTIR spectra provide strong evidence for retained PVP in
S2--S4 through the combined C=O, C--N, and C--H vibrational
fingerprint, with the band near 1428~cm$^{-1}$ additionally assigned
in the literature to coordination between PVP nitrogen and Zn$^{2+}$
ions \cite{ankinapalli2024synthesis}. The low-frequency absorption is
compatible with ZnS-containing nanoparticle powders but is not
sufficiently specific to establish the crystal structure. The XRD
analysis that follows is therefore used to verify whether the cubic
ZnS phase is retained and to determine whether increasing PVP loading
affects the diffraction-line widths, peak positions, or coherent-domain
sizes.

\subsection{SEM and EDS Analysis}
\begin{figure}[ht!]
\centering
\includegraphics[width=0.8\textwidth]{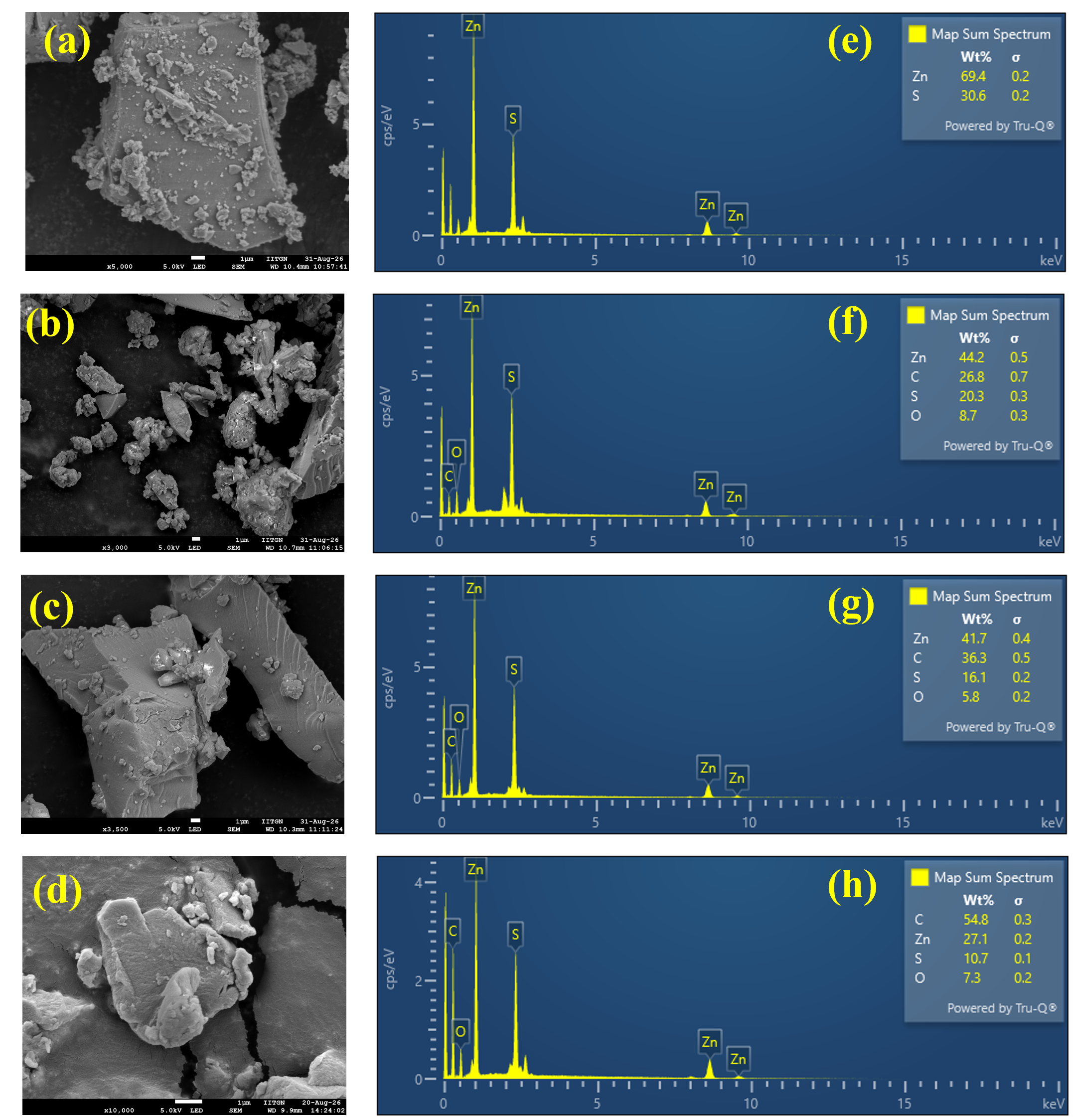}
\caption{SEM micrographs and corresponding EDS spectra of pristine and PVP-modified ZnS nanoparticles. The left column shows the SEM micrographs of (a) S1, (b) S2, (c) S3, and (d) S4, while the right column shows the corresponding EDS spectra of (e) S1, (f) S2, (g) S3, and (h) S4.}
\label{fig:sem_eds}
\end{figure}
Figure~\ref{fig:sem_eds} shows the SEM micrographs and EDS spectra of pristine and PVP-modified ZnS nanoparticles. The SEM micrographs are presented in the left column, with panels (a), (b), (c), and (d) corresponding to S1, S2, S3, and S4, respectively. Their corresponding EDS spectra are displayed in the right column as panels (e), (f), (g), and (h). The SEM images reveal that all four samples possess agglomerated morphologies, with noticeable variations in the shape, packing, and surface texture of the structures as the PVP content is increased.

The pristine ZnS sample S1, shown in Fig.~\ref{fig:sem_eds}(a), consists predominantly of large, compact, block-like aggregates. Numerous smaller grains are distributed over the surfaces of these structures, producing a rough and heterogeneous surface texture. Such agglomeration is commonly observed in chemically synthesized ZnS nanoparticles because of strong interactions between neighbouring particles \cite{Koczkur2015PVP,Poornaprakash2017}.

The morphology changes appreciably after the introduction of PVP. Sample S2 in Fig.~\ref{fig:sem_eds}(b) contains comparatively dispersed and fragmented clusters with irregular shapes and a broad distribution of aggregate sizes. The reduced compactness of these structures indicates that PVP influences the association and organization of ZnS particles during precipitation. In S3, shown in Fig.~\ref{fig:sem_eds}(c), broader plate-like structures become more prominent. These plates contain smaller particles attached to their surfaces and edges, giving rise to a morphology composed of extended structures together with granular deposits. At the highest PVP content, S4 in Fig.~\ref{fig:sem_eds}(d) exhibits compact flake-like and layered structures accompanied by finer surface grains.

The observed morphological changes show that PVP affects the nucleation, growth, and assembly of the ZnS particles. The polar pyrrolidone groups of PVP can interact with Zn sites on the developing ZnS surfaces, while the polymer chains influence the contact and association between neighbouring particles. Adsorption of PVP consequently modifies the surface energy and relative growth of different ZnS surfaces, producing the distinct aggregate morphologies observed for S2--S4 \cite{Ghosh2006PVPZnS,Soltani2012PVPZnS,Koczkur2015PVP}. Similar changes in the morphology and aggregation of PVP-modified ZnS nanoparticles have been reported for materials prepared through wet-chemical synthesis \cite{Poornaprakash2017,Kumar2022PVPCapped}. The observed morphological changes are consistent with an influence of
PVP on particle association and growth during the precipitation process.

The EDS spectrum of pristine S1 is shown in
Fig.~\ref{fig:sem_eds}(e). Strong Zn and S signals are observed, with
reported contents of 69.4~wt.\,\% and 30.6~wt.\,\%, respectively. The
spectrum contains the characteristic low-energy Zn $L$ signal, the S
$K$ signal near 2.3~keV, and the higher-energy Zn $K$ signals. These
features establish the presence of Zn and S in the pristine sample.
The EDS spectrum supports the assignment of S1 as ZnS. EDS
alone identifies the constituent elements but does not determine the
crystalline phase.

The EDS spectra of the PVP-modified samples S2--S4 are shown in Fig.~\ref{fig:sem_eds}(f)--(h), respectively. All three spectra show Zn and S together with clear carbon and oxygen signals. The reported carbon fraction increases from 26.8~wt.\,\% in S2 to 54.8~wt.\,\% in S4, accompanied by decreases in the relative Zn and S fractions. Oxygen is detected in all three samples but shows no monotonic trend. These observations are consistent with the presence of polymer-derived material in the analyzed regions, although the EDS fractions alone cannot quantify retained PVP loading or surface coverage.

Carbon and oxygen are constituent elements of the pyrrolidone repeat
unit of PVP. Their appearance in S2--S4, together with the
PVP-related vibrational bands identified by FTIR, supports the
retention of PVP-derived material in the modified ZnS samples. The
reported carbon contribution increases systematically from
26.8~wt.\,\% for S2 to 36.3~wt.\,\% for S3 and
54.8~wt.\,\% for S4. This progression follows the increasing nominal
amount of PVP used during synthesis and is consistent with a larger
contribution from carbon-rich PVP-derived material across the series.
The EDS weight fractions should not, however, be interpreted as a
direct quantitative measurement of the amount of PVP attached to the
ZnS particles. The EDS signal represents the composition of the
region sampled by the electron beam and depends on the measurement
and sample-preparation conditions.

The oxygen signal is also consistent with the presence of PVP because
each pyrrolidone repeat unit contains a carbonyl group. However, EDS
cannot identify a particular functional group or establish bonding
between the carbonyl oxygen and surface Zn atoms. Evidence for the
chemical environment of the carbonyl group is instead provided by the
FTIR spectrum. The observed PVP-related FTIR bands, including the
carbonyl vibration, are consistent with an interaction between PVP
and the ZnS particle surface. The continued presence of Zn and S in S2--S4 confirms that
the inorganic ZnS component is retained after modification, while the
appearance of carbon and oxygen is consistent with the addition of
PVP-derived material.

The EDS and FTIR results therefore provide complementary information.
EDS establishes the presence of Zn and S in all four samples and
shows an increasing carbon-rich contribution across S2--S4. FTIR
identifies the characteristic C--N, C=O, CH$_2$, and C--H vibrations
associated with PVP. When considered together with the SEM
observations, these results support the association of PVP-derived
material with the ZnS particles and show that the modified preparation
conditions influence their morphology and surface composition.

\subsection{Crystal Structure and Crystallite-Size Analysis}
\label{subsec:xrd_analysis}

\begin{figure}[htbp]
    \centering
    \includegraphics[width=\linewidth]{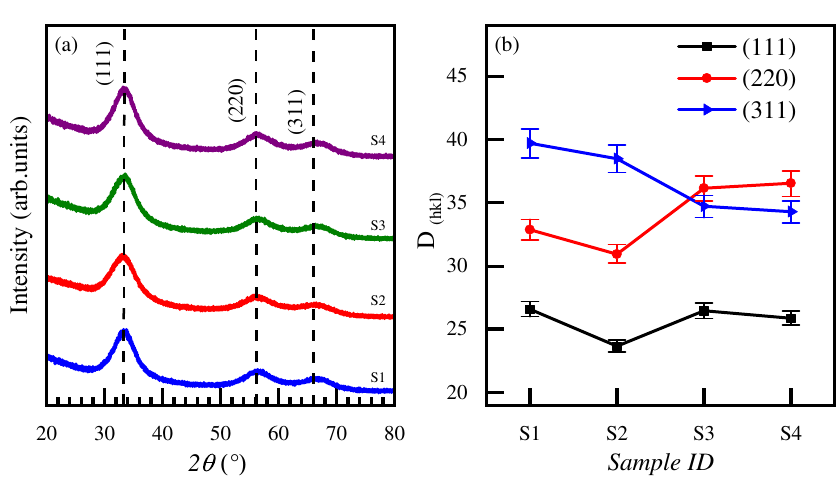}
    \caption{(a) Room-temperature X-ray diffraction patterns of pristine ZnS (S1) and PVP-modified ZnS nanoparticles (S2--S4), recorded using Co $K_{\alpha}$ radiation. The dashed vertical lines indicate the (111), (220), and (311) reflections of cubic ZnS. The patterns are vertically displaced for clarity. (b) Apparent Scherrer coherent-domain sizes, $D_{hkl}$, obtained from the three reflections and expressed in \AA.}
    \label{fig:xrd_analysis}
\end{figure}

X-ray diffraction (XRD) was used to examine the crystalline phase, diffraction-peak positions, and apparent coherent-domain sizes of pristine ZnS (S1) and PVP-modified ZnS nanoparticles (S2--S4). Figure~\ref{fig:xrd_analysis}(a) presents the room-temperature XRD patterns, while Fig.~\ref{fig:xrd_analysis}(b) shows the apparent Scherrer coherent-domain sizes obtained from the three principal reflections.

As shown in Fig.~\ref{fig:xrd_analysis}(a), all four samples exhibit broad diffraction maxima corresponding to the (111), (220), and (311) planes of cubic zinc-blende ZnS, with space group $F\bar{4}3m$. These reflections are consistent with the standard powder-diffraction reference for cubic ZnS (PDF/JCPDS No.~05-0566) \cite{VigneshwariFeZnS2026,VigneshwariCoZnS2026,Ghosh2006PVPZnS,Soltani2012PVPZnS,He2003}. Their retention across S1--S4 indicates that the principal crystalline phase is preserved after PVP-assisted preparation.

No additional well-resolved reflections attributable to crystalline secondary phases are observed in Fig.~\ref{fig:xrd_analysis}(a) within the measured angular range. Retained PVP may contribute diffuse scattering because of its predominantly amorphous character \cite{Ghosh2006PVPZnS,Soltani2012PVPZnS,Koczkur2015PVP,He2003}. These observations do not exclude amorphous material, minor secondary phases, or poorly crystalline components below the detection capability of the measurement.

The diffraction patterns in Fig.~\ref{fig:xrd_analysis}(a) show small variations in peak position across S1--S4. Within the PVP-modified series, the reflections move slightly toward higher angles from S2 to S4, while their positions relative to pristine S1 depend on the reflection considered. The changes are modest and do not indicate a systematic displacement of all reflections in the same direction.

According to Bragg's law, a lower diffraction angle corresponds to a larger interplanar spacing, whereas a higher angle corresponds to a smaller spacing. However, the reflection-dependent changes observed here do not establish uniform expansion or contraction of the cubic lattice. Residual strain, surface stress, structural disorder, and experimental peak-position uncertainty may contribute to the differences \cite{Soltani2012PVPZnS,WilliamsonHall1953,Balzar2004}. PVP can modify the growth and surface environment of ZnS during precipitation, but these peak shifts do not independently identify a particular strain or defect mechanism \cite{Ghosh2006PVPZnS,Soltani2012PVPZnS,Koczkur2015PVP,He2003}.

The broad diffraction maxima in Fig.~\ref{fig:xrd_analysis}(a) are consistent with nanometre-scale coherent domains, although finite size is not the only source of line broadening. Instrumental resolution, lattice microstrain, and structural defects may also contribute \cite{WilliamsonHall1953,Balzar2004,LangfordWilson1978}. The apparent coherent-domain size normal to each set of planes was estimated using the Scherrer equation \cite{LangfordWilson1978,Dsingh2024skew,HolzwarthGibson2011,Hargreaves2016},
\begin{equation}
D_{hkl}
=
\frac{K\lambda_{\mathrm{Co}}}
{\beta_{hkl}\cos\theta_{hkl}},
\label{eq:scherrer_xrd}
\end{equation}
where $K$ is the crystallite-shape factor, $\lambda_{\mathrm{Co}}=1.7906~\text{\AA}$ is the Co $K_{\alpha}$ wavelength, $\beta_{hkl}$ is the full width at half maximum of the diffraction peak expressed in radians of $2\theta$, and $\theta_{hkl}$ is the Bragg angle. The shape factor depends on crystallite morphology and the adopted line-width definition \cite{Singh2022deadlayerstatic}. A value of $K=0.91$ was used for the present calculations \cite{singh2020}.

The calculated values are treated as apparent Scherrer coherent-domain sizes rather than direct particle diameters because size-related and other broadening contributions are not independently separated \cite{Balzar2004,LangfordWilson1978,Hargreaves2016}. The reflection-dependent sizes are plotted in Fig.~\ref{fig:xrd_analysis}(b) and listed in Table~\ref{tab:scherrer_sizes}.

\begin{table}[htbp]
\centering
\caption{Apparent Scherrer coherent-domain sizes obtained from the (111), (220), and (311) reflections of pristine and PVP-modified ZnS nanoparticles. The final column gives the arithmetic mean of the three reflection-dependent values, calculated from the unrounded results.}
\label{tab:scherrer_sizes}
\begin{tabular}{
>{\centering\arraybackslash}m{1.2cm}
>{\centering\arraybackslash}m{2cm}
>{\centering\arraybackslash}m{2cm}
>{\centering\arraybackslash}m{2cm}
>{\centering\arraybackslash}m{2cm}}
\toprule
Sample &
$D_{111}$ (\AA) &
$D_{220}$ (\AA) &
$D_{311}$ (\AA) &
$\overline{D}$ (nm) \\
\midrule
S1 & 26.58 & 32.88 & 39.70 & 3.31 \\
S2 & 23.68 & 30.94 & 38.49 & 3.10 \\
S3 & 26.47 & 36.14 & 34.71 & 3.24 \\
S4 & 25.86 & 36.53 & 34.28 & 3.22 \\
\bottomrule
\end{tabular}
\end{table}

The reflection-dependent sizes shown in Fig.~\ref{fig:xrd_analysis}(b) and listed in Table~\ref{tab:scherrer_sizes} span 23.68--39.70~\AA, corresponding to approximately 2.37--3.97~nm. For the (111) reflection, the apparent size decreases from 26.58~\AA{} in S1 to 23.68~\AA{} in S2, increases to 26.47~\AA{} in S3, and decreases to 25.86~\AA{} in S4. The (220) size also initially decreases, from 32.88~\AA{} in S1 to 30.94~\AA{} in S2, before increasing to 36.14 and 36.53~\AA{} in S3 and S4, respectively. In contrast, the central values of $D_{311}$ decrease throughout the series, from 39.70~\AA{} in S1 to 38.49, 34.71, and 34.28~\AA{} in S2--S4.

Figure~\ref{fig:xrd_analysis}(b) also shows that the (311) reflection gives the largest apparent size in S1 and S2, whereas the (220) reflection gives the largest value in S3 and S4. These differences show that the reflection-dependent broadening does not follow a single uniform trend. Direction-dependent domain dimensions, strain, defects, and peak-width uncertainty can all affect the calculated values. The results therefore do not independently demonstrate anisotropic particle growth or establish that every difference between samples is statistically significant.

The arithmetic mean sizes listed in Table~\ref{tab:scherrer_sizes} decrease from 3.31~nm for S1 to 3.10~nm for S2, increase to 3.24~nm for S3, and decrease slightly to 3.22~nm for S4. Thus, S2 has the smallest mean apparent coherent-domain size, and the mean varies non-monotonically with PVP addition. All four values remain within the narrow interval of 3.10--3.31~nm. These arithmetic means provide a descriptive summary of the three reflections rather than an independently determined particle-size distribution.

PVP adsorption during precipitation can modify precursor attachment, particle association, and surface relaxation \cite{Ghosh2006PVPZnS,Soltani2012PVPZnS,Koczkur2015PVP,He2003}. Such effects provide possible explanations for the observed differences, but the present diffraction analysis cannot distinguish changes in domain dimensions from changes in strain- or defect-related broadening.

Overall, the diffraction patterns in Fig.~\ref{fig:xrd_analysis}(a) support retention of the principal cubic zinc-blende ZnS phase throughout S1--S4. The sizes in Fig.~\ref{fig:xrd_analysis}(b) and Table~\ref{tab:scherrer_sizes} show small, reflection-dependent variations, while the mean Scherrer size remains close to 3~nm. The results do not establish a monotonic reduction in coherent-domain size with increasing PVP addition.

\subsection{Transmission electron microscopy} 

The XRD results establish that S1 and S4 retain the cubic ZnS phase, while the SEM images show that both powders form much larger aggregates. To further examine the relationship between aggregate morphology and local crystalline structure, transmission electron microscopy (TEM) was performed on the two end members of the series, pristine S1 and the highest-PVP capped ZnS sample S4.

For grid preparation, 1 mg of each powder was dispersed in 1 mL of deionized water. A 50~\(\mu\mathrm{L}\) aliquot of this dispersion was diluted with deionized water to a final volume of 1 mL, giving a nominal concentration of 0.05 mg mL\(^{-1}\) before sonication. The diluted preparation was sonicated for 30 min, after which the sample was deposited onto a 200-mesh carbon-coated copper TEM grid and dried under ambient conditions. The grid-deposition and drying steps follow related ZnS preparations~\cite{VigneshwariFeZnS2026,VigneshwariCoZnS2026}. TEM, high-resolution TEM (HRTEM), and selected-area electron diffraction (SAED) were performed using the 300 kV instrument described in Sec.~2.4.

\begin{figure}[!htbp]
  \centering
  \includegraphics[width=\linewidth]{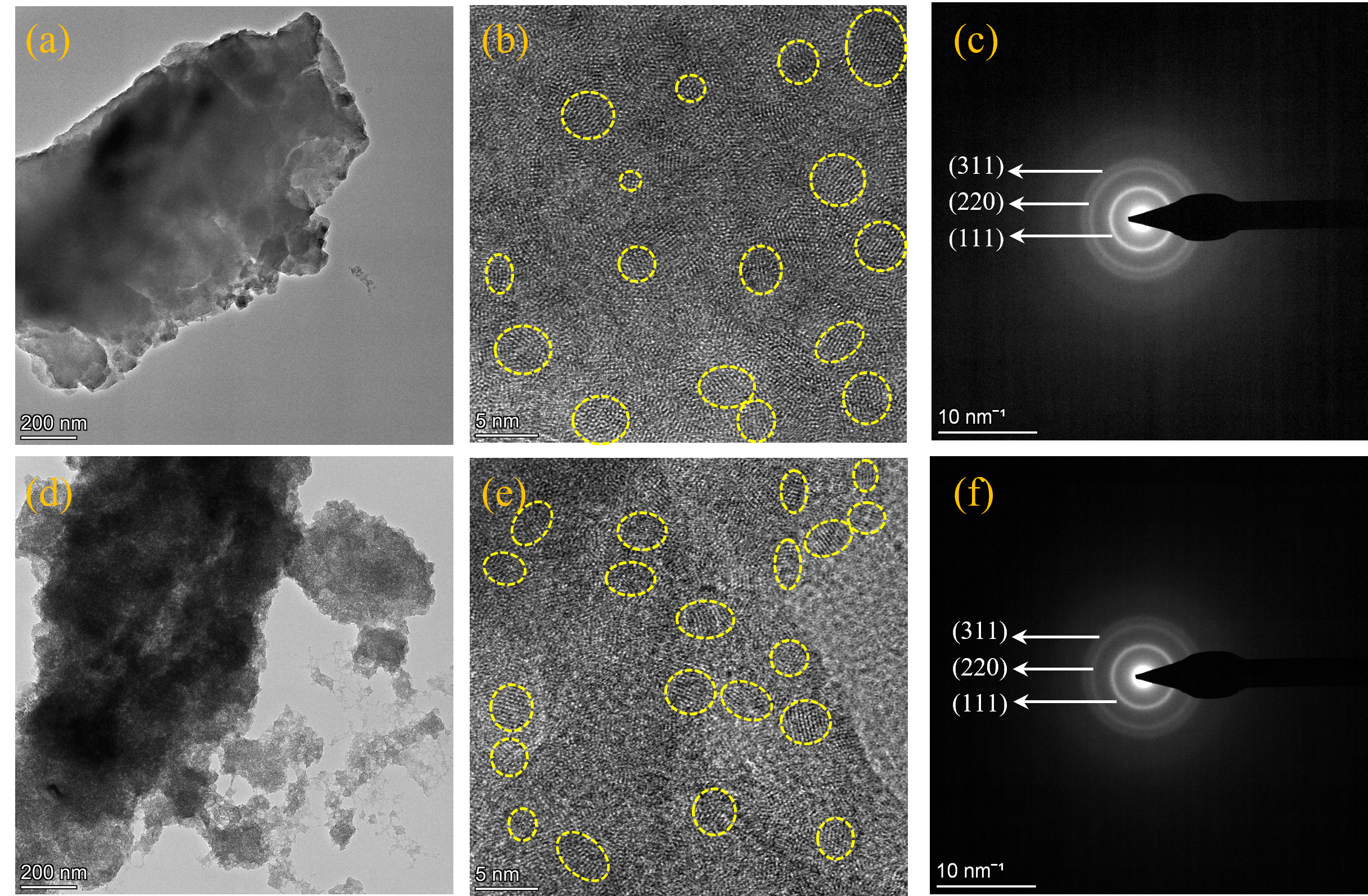}
  \caption{TEM analysis of pristine ZnS (S1) and the highest-PVP ZnS sample (S4). Panels (a--c) correspond to S1 and panels (d--f) to S4. Panels (a,d) show lower-magnification TEM images, (b,e) show HRTEM images, and (c,f) show SAED patterns.}
  \label{fig:tem_microscopy}
\end{figure}

The lower-magnification images in Fig.~\ref{fig:tem_microscopy}(a,d) show aggregates in both samples. S1 contains a large, dense, irregular region with finer features near its edges, whereas S4 shows a dense region accompanied by smaller clusters. This qualitative difference is consistent with the SEM observations of large block-like aggregates in S1 and finer surface features accompanying the flake-like structures in S4. TEM provides a closer view of selected regions, however, and both samples remain agglomerated in the displayed fields. These images therefore support a difference in aggregate morphology without establishing a reduction in the overall degree of agglomeration or an aggregate-size distribution for S4. PVP-dependent changes in ZnS growth and dispersion have been reported under other synthesis conditions~\cite{Ghosh2006PVPZnS,Soltani2012PVPZnS}.

The HRTEM images in Fig.~\ref{fig:tem_microscopy}(b,e) reveal locally resolved lattice fringes on the nanometre scale in both S1 and S4. The presence of differently oriented lattice fringes within the imaged regions is consistent with multiple crystalline domains within the larger aggregates. Thus, HRTEM confirms the presence of locally ordered crystalline regions within the aggregated structures observed in the lower-magnification TEM images.

The SAED patterns in Fig.~\ref{fig:tem_microscopy}(c,f) contain diffraction rings that can be indexed to the (111), (220), and (311) planes of cubic zinc-blende ZnS. These correspond to the principal reflections observed in the powder XRD patterns in Fig.~\ref{fig:xrd_analysis}. The ring-like nature of the SAED patterns is characteristic of polycrystalline samples, arising from crystalline domains with different orientations within the selected areas~\cite{Maity2026657,Dsingh2025viable,Maity20261077}. The agreement between the SAED and XRD indexing therefore supports the retention of the cubic ZnS structure in both S1 and S4, consistent with earlier reports~\cite{VigneshwariCoZnS2026,VigneshwariFeZnS2026}. Similar correspondence between XRD and SAED has been reported for polymer-encapsulated ZnS nanoparticles~\cite{Devi2020ZnSCMC}. Neither the SAED patterns nor the HRTEM images directly identify a PVP surface layer.

\begin{figure}[!htbp]
  \centering
  \includegraphics[width=\linewidth]{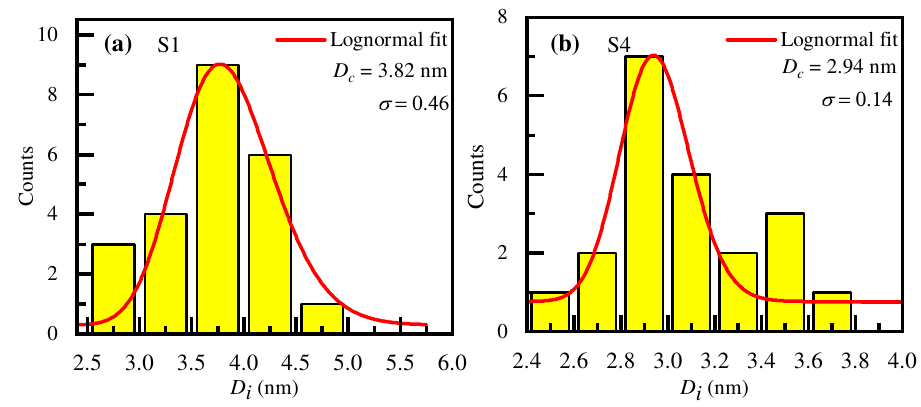}
  \caption{TEM-derived crystallite-size distributions for (a) pristine ZnS (S1) and (b) the highest-PVP ZnS sample (S4), with the corresponding log-normal fits.}
  \label{fig:tem_size}
\end{figure}

To describe the crystallite dimensions resolved in the TEM images, the diameters of locally distinguishable crystalline regions were measured for S1 and S4 and plotted as size-distribution histograms. The distributions were fitted using a log-normal function, which is commonly used to describe size distributions in nanoscale and granular systems~\cite{Knobel2008,Bedanta2009}. The distribution was expressed as
\begin{equation}
f(D_i)=
\frac{1}{\sqrt{2\pi}\,\sigma D_i}
\exp\left[
-\frac{\left[\ln(D_i/D_c)\right]^2}{2\sigma^2}
\right],
\label{eq:lognormal_distribution}
\end{equation}
where \(D_i\) is the measured crystallite diameter, \(D_c\) is the median or characteristic crystallite diameter, and \(\sigma\) is the standard deviation of \(\ln(D_i)\). The measured distributions and corresponding log-normal fits are shown in Fig.~\ref{fig:tem_size}. For S1, the fitted characteristic crystallite diameter is \(D_c=3.82\)~nm with \(\sigma=0.46\), whereas S4 gives \(D_c=2.94\)~nm with \(\sigma=0.14\). Thus, the characteristic crystallite dimension obtained from the selected TEM regions decreases from S1 to S4. The smaller value of \(\sigma\) for S4 indicates a narrower fitted distribution of the measured crystallite dimensions.

The TEM observations provide local structural context for the morphology observed by SEM and the coherent-domain sizes estimated from XRD. The TEM images show comparatively larger structural features for S1 than for S4, while the fitted characteristic crystallite diameter decreases from \(3.82\)~nm for S1 to \(2.94\)~nm for S4. The arithmetic mean of the reflection-dependent Scherrer sizes is also slightly larger for S1 (3.31~nm) than for S4 (3.22~nm). The HRTEM images show local crystalline order within the aggregates, and the SAED ring assignments support the cubic ZnS phase identified by XRD.

\subsection{UV--Visible Absorption and Optical Band-Gap Analysis}
\label{subsec:uv_visible}

\begin{figure*}[htbp]
    \centering
    \includegraphics[width=1.0\textwidth]{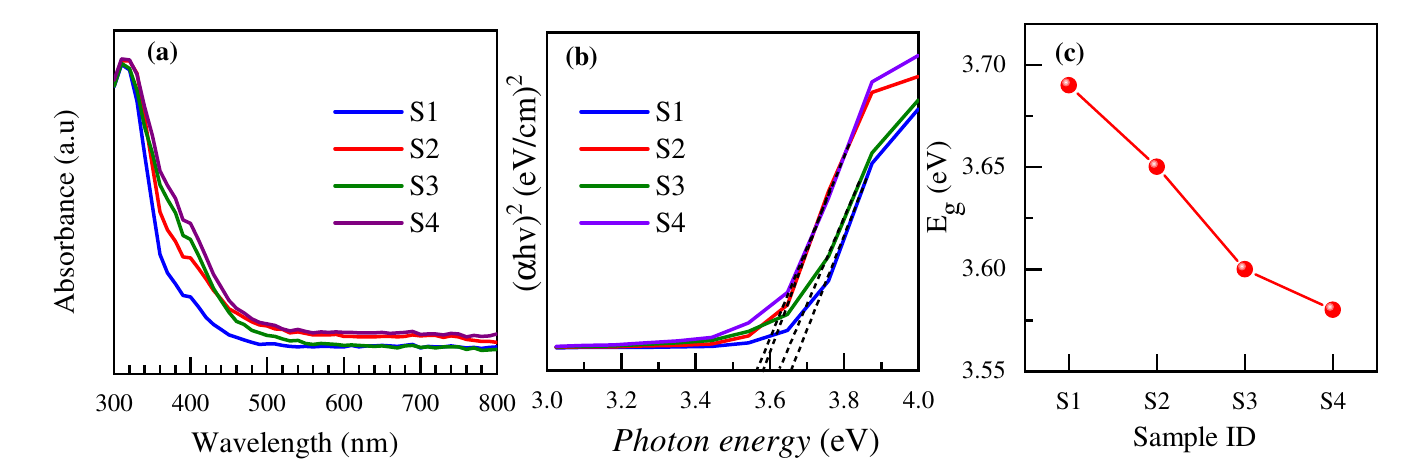}
    \caption{(a) Room-temperature UV--visible absorbance spectra of pristine
    ZnS (S1) and PVP-modified ZnS nanoparticles (S2--S4). (b) Corresponding
    allowed-direct-transition Tauc plots. The dashed lines indicate the linear
    regions extrapolated to the photon-energy axis. (c) Variation of the
    extracted Tauc optical gap with sample ID.}
    \label{fig:absor_spectra}
\end{figure*}

The FTIR and XRD results provide the structural basis for interpreting the
optical response. FTIR indicates that the pyrrolidone groups of PVP interact
with the ZnS nanoparticle surface, whereas XRD confirms retention of the cubic
ZnS phase throughout the S1--S4 series without a detectable crystalline
secondary phase. The changes in the optical spectra can therefore be discussed
within a structurally preserved ZnS framework, while recognizing that PVP can
modify precursor coordination, nanoparticle growth, aggregation, surface
passivation, and the local defect environment
\cite{Ghosh2006PVPZnS,Soltani2012PVPZnS,Koczkur2015PVP,Fang2011}.

Figure~\ref{fig:absor_spectra}(a) shows strong ultraviolet absorption for all
four samples, followed by a rapid decrease in absorbance toward the visible
region. This response is characteristic of wide-band-gap ZnS nanostructures
\cite{Nanda2000,Fang2011,Ghosh2006PVPZnS,Shahi2017}. The absorption edge read from
the spectra occurs at approximately 310~nm for pristine ZnS (S1) and at about
320~nm for the PVP-modified samples S2--S4. Thus, the first addition of PVP
moves the absorption onset toward longer wavelength, after which the graphical
edge position remains close to 320~nm over the investigated PVP range.

A displacement of an absorption feature toward longer wavelength is termed a
\emph{red shift}. Because photon energy and wavelength are inversely related,
\begin{equation}
E(\mathrm{eV})
= \frac{hc}{\lambda}
\simeq \frac{1239.84}{\lambda(\mathrm{nm})}.
\end{equation}
the change from 310 to 320~nm corresponds to a reduction in the associated
photon energy from approximately 4.00 to 3.88~eV. The observed red shift
therefore means that the PVP-modified samples begin to absorb at a lower photon
energy than pristine S1 \cite{Nanda2000,Fang2011,Makula2018}.

In addition to the change in edge position, the absorption profile becomes
broader and extends farther toward longer wavelength with increasing PVP
content. Absorption-edge broadening in semiconductor nanoparticles may arise
from a distribution of particle environments, local strain, structural
disorder, surface coordination, stoichiometric defects, and localized states
near the band edges \cite{Fang2011,Koczkur2015PVP,Prasad2019}. The present spectra
are therefore consistent with a progressively modified ZnS--PVP interfacial
environment.

ZnS is generally treated as a direct-band-gap semiconductor. Accordingly, the
optical gap was evaluated using the allowed-direct-transition form of the Tauc
relation \cite{Tauc1966,Nanda2000,Fang2011,Makula2018},
\begin{equation}
(\alpha h\nu)^2=B\left(h\nu-E_{\mathrm{g}}\right),
\label{eq:tauc_direct_zns}
\end{equation}
where $\alpha$ is the absorption coefficient, $h\nu$ is the incident photon
energy, $B$ is a proportionality constant, and $E_{\mathrm{g}}$ is the Tauc
optical gap. When all samples are recorded using the same optical path length,
absorbance is proportional to $\alpha$; multiplication by a constant changes
the vertical scale of the Tauc plot but not the photon-energy intercept
\cite{Tauc1966,Makula2018}. For each sample, the approximately linear part of
the $(\alpha h\nu)^2$ versus $h\nu$ curve was fitted and extrapolated to
$(\alpha h\nu)^2=0$, as shown in Fig.~\ref{fig:absor_spectra}(b).

The extracted values are summarized graphically in
Fig.~\ref{fig:absor_spectra}(c). S1 has
the largest Tauc optical gap, $E_{\mathrm{g}}=3.69$~eV. The gap decreases to
3.65~eV for S2, 3.60~eV for S3, and 3.58~eV for S4, corresponding to reductions
of 0.04, 0.09, and 0.11~eV relative to pristine ZnS. The S1 value is close to
the commonly reported room-temperature gap of cubic ZnS
\cite{Nanda2000,Fang2011,Soltani2012PVPZnS}. The monotonic decrease across S1--S4 is
consistent with the increasing contribution of lower-energy near-edge
absorption in the PVP-containing samples.

PVP is not expected to act as a conventional
substitutional dopant in these samples; rather, its carbonyl-containing groups
coordinate at the nanoparticle surface and alter the growth and interfacial
environment \cite{Ghosh2006PVPZnS,Soltani2012PVPZnS,Koczkur2015PVP}. PVP-associated changes
in particle-size distribution, surface strain, passivation, and the population
of surface or defect-related states can broaden the near-edge response and
shift the Tauc intercept without requiring formation of a new crystalline
phase \cite{Ghosh2006PVPZnS,Shahi2017,Prasad2019,Fang2011}. PVP-related electronic
transitions and energy-transfer effects have also been reported in capped ZnS
systems, further demonstrating that the polymer shell can participate in the
optical excitation environment \cite{Manzoor2004}.

The smaller gap change between S3 and S4 (0.02~eV) than between S2 and S3
(0.05~eV) indicates a leveling of the extracted Tauc-gap trend at the highest
PVP loading. This observation may reflect a smaller incremental modification
of the accessible surface environment, but it is not sufficient by itself to
establish saturation of PVP coverage\cite{Koczkur2015PVP,Prasad2019,Makula2018}.

Overall, PVP modification produces a red shift of the measured absorption edge
from approximately 310 to 320~nm and a monotonic reduction of the Tauc optical
gap from 3.69 to 3.58~eV. Together with the FTIR evidence for ZnS--PVP surface
interaction and the XRD evidence for retention of cubic ZnS, the optical data
support an interpretation based on modification of the nanoparticle growth,
surface coordination, strain/disorder, and localized near-edge states rather
than formation of a new narrow-gap crystalline phase
\cite{Ghosh2006PVPZnS,Soltani2012PVPZnS,Koczkur2015PVP,Fang2011,Prasad2019}.

\subsection{Temperature-Dependent Raman Spectroscopy}
\label{subsec:temperature_raman}

Raman spectroscopy was employed to investigate the vibrational modes and structural characteristics of pristine and PVP-modified ZnS nanoparticles, following approaches reported in the literature for thin films and nanoparticles \cite{DSingh2025Raman, TVV202641}. The FTIR, SEM-EDS, XRD, and UV--visible measurements provide the structural and optical context for the Raman analysis. XRD shows that all four samples retain the cubic zinc-blende ZnS phase. The Tauc optical gap  decreases from $3.69~\mathrm{eV}$ for pristine ZnS (S1) to $3.58~\mathrm{eV}$ for the
highest-PVP sample (S4). Temperature-dependent Raman spectroscopy was therefore
 used to examine whether this optical evolution is accompanied by a
systematic modification of the ZnS lattice vibrations. Spectra were recorded
from 100 to 300~K in intervals of 20~K.

\begin{figure*}[t]
\centering
\includegraphics[width=\textwidth]
{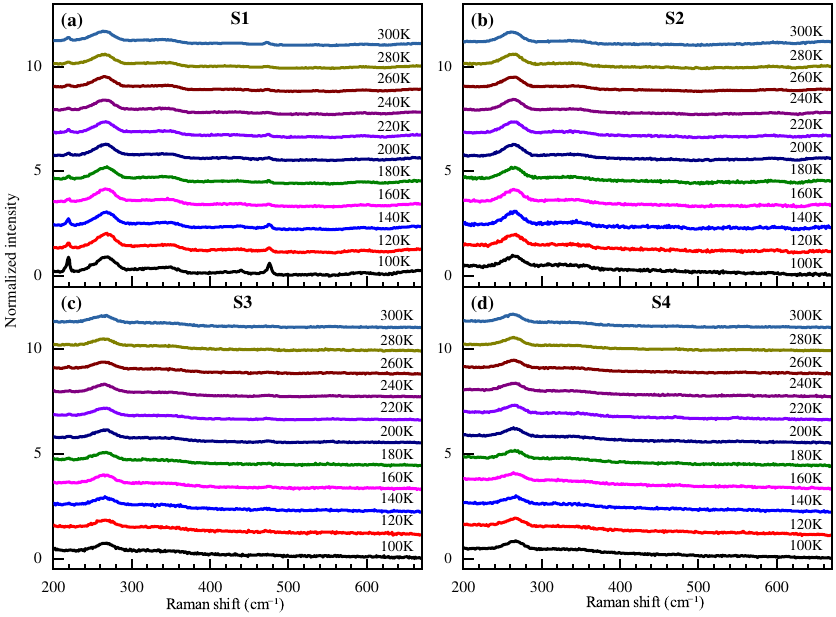}
\caption{Temperature-dependent Raman spectra of
(a) pristine ZnS (S1), (b) low-PVP ZnS (S2),
(c) intermediate-PVP ZnS (S3), and (d) highest-PVP ZnS (S4),
recorded from 100 to 300~K in intervals of 20~K. Each spectrum was
independently normalized and vertically displaced for clarity.}
\label{fig:raman_temperature_spectra}
\end{figure*}

Figure~\ref{fig:raman_temperature_spectra} presents the complete temperature
series. Every sample exhibits a broad dominant band near
$261- 267~\mathrm{cm}^{-1}$ and a weaker contribution in the
$340$--$355~\mathrm{cm}^{-1}$ region. In all four samples, the dominant band
shifts to higher wavenumber on cooling, while its broad profile is retained
throughout the measured temperature range. No new dominant Raman band or
abrupt spectral change appears between 100 and 300~K. Thus, PVP modification
does not remove the characteristic ZnS vibrational response, and there is no
Raman evidence of a temperature-induced structural transition over the
investigated range.

The dominant band occurs between $261.84$ and $267.00~\mathrm{cm}^{-1}$
across the complete dataset and is assigned to a TO-like ZnS phonon. In bulk
cubic ZnS, the transverse-optical (TO) and longitudinal-optical (LO) modes
occur near $270$--$273$ and $350$--$352~\mathrm{cm}^{-1}$, respectively
\cite{Nilsen1969,Cheng2009}. The lower frequency and broad profile of
the present TO-like band are consistent with the nanocrystalline character of
the samples and a distribution of local vibrational environments. Finite
crystallite size, local strain, structural disorder, surface bonding, and the
crystallite-size distribution can all contribute to the observed shift and
broadening. 

The weaker LO-like feature has median positions of approximately
$349.8$, $346.3$, $352.0$, and $346.4~\mathrm{cm}^{-1}$ for S1, S2, S3,
and S4, respectively. Its peak height is only approximately 11--20\% of the TO-like band. This feature supports the presence of the expected ZnS
optical-mode region but is too weak and broad for a reliable independent
temperature-dependent fit. The quantitative temperature analysis is therefore
restricted to the stronger TO-like band.

The TO-like band was fitted over $235$--$295~\mathrm{cm}^{-1}$ using a
Gaussian profile together with a local background. The fitted centre was
described by the linear relation
\begin{equation}
    \widetilde{\nu}_{\mathrm{TO}}(T)
    =
    \widetilde{\nu}_{\mathrm{int}}+\chi T,
    \label{eq:raman_linear_temperature}
\end{equation}
where $\widetilde{\nu}_{\mathrm{int}}$ is the fitted intercept and
$\chi=d\widetilde{\nu}_{\mathrm{TO}}/dT$ is the first-order temperature
coefficient. This linear relation is commonly used to determine Raman
temperature coefficients over a restricted temperature interval
\cite{Singh2021RamanThermometry}. Over the measured range, the linear model
captures the dominant temperature dependence. The fitted coefficients
are summarized in Table~\ref{tab:raman_temperature_coefficients}.

\begin{table*}[t]
\centering
\caption{Temperature dependence of the main TO-like Raman band. The
cooling-induced phonon hardening is defined as
$\Delta\widetilde{\nu} =
\widetilde{\nu}(100~\mathrm{K}) -
\widetilde{\nu}(300~\mathrm{K})$.
Uncertainties in $\chi$ are the standard errors of the linear fits.}
\label{tab:raman_temperature_coefficients}
\small

\begin{tabular}{c c c c c c}
\hline
Sample &
$\widetilde{\nu}(100~\mathrm{K})$ &
$\widetilde{\nu}(300~\mathrm{K})$ &
$\Delta\widetilde{\nu}$ &
$\chi$ &
$R^2$ \\
&
($\mathrm{cm}^{-1}$) &
($\mathrm{cm}^{-1}$) &
($\mathrm{cm}^{-1}$) &
($\mathrm{cm}^{-1}\,\mathrm{K}^{-1}$) &
\\
\hline

S1 & 266.87 & 263.68 & 3.19 & $-0.0166 \pm 0.0011$ & 0.964 \\

S2 & 264.11 & 261.84 & 2.27 & $-0.0102 \pm 0.0015$ & 0.830 \\

S3 & 265.94 & 263.78 & 2.16 & $-0.0103 \pm 0.0009$ & 0.930 \\

S4 & 265.51 & 262.80 & 2.71 & $-0.0137 \pm 0.0016$ & 0.887 \\

\hline
\end{tabular}

\end{table*}
\begin{figure*}[t]
\centering
\includegraphics[width=0.6\textwidth]
{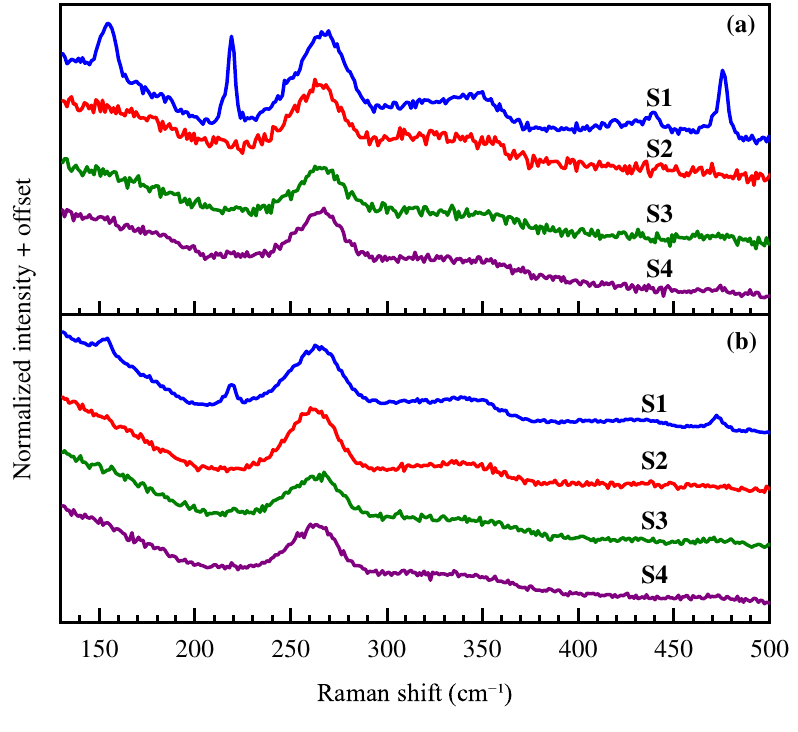}
\caption{Normalized, vertically offset Raman spectra of S1–S4 at 100~K (a) and 300~K (b).}
\label{fig:raman_quantitative_analysis}
\end{figure*}

The endpoint spectra in Fig.~\ref{fig:raman_quantitative_analysis}(a) and
(b) provide a direct visual comparison of the Raman response at 100 and
300~K. At 100~K, the TO-like maxima occur at higher wavenumbers than their
corresponding positions at 300~K, in agreement with the quantitative shifts
in Table~\ref{tab:raman_temperature_coefficients}. Apart from this displacement,
the broad TO-like profile and the weak LO-like region remain recognizable at
both temperatures. No additional dominant band appears on cooling, and none
of the principal features disappears on heating. The comparison therefore
shows that the main temperature-dependent change is a continuous shift of the
TO-like band rather than a qualitative reconstruction of the Raman spectrum.

Because the spectra in Fig.~\ref{fig:raman_quantitative_analysis}(a) and (b) were normalized
independently, these panels are used only to compare band positions,
overall line shapes, and the presence or absence of weak spectral
features. In particular, they do not provide a basis for extracting
temperature-dependent TO or LO intensities.

Additional weak bands provide information about the sulfur environment.
S1 contains correlated features near $153.2$, $219.0$, and
$474.2~\mathrm{cm}^{-1}$. These features are particularly clear in the
100~K endpoint spectrum in Fig.~\ref{fig:raman_quantitative_analysis}(a)
and remain identifiable, although less sharply resolved, in the 300~K
spectrum in Fig.~\ref{fig:raman_quantitative_analysis}(b).
Elemental sulfur has characteristic Raman bands near $153$, $219$,
and $472$--$474~\mathrm{cm}^{-1}$, associated with bending and stretching
vibrations of the S--S structure \cite{Nims2019Sulfur}. The correlated
presence of all three features is therefore consistent with a small
contribution from residual elemental sulfur in S1. Similar features are
weaker or poorly resolved in S2, S3 and S4. This
difference may indicate a smaller Raman-visible contribution from residual
sulfur or a change in its surface environment after PVP addition.

Taken together, the Raman results show that the TO-like band undergoes continuous hardening on cooling in all four samples. The shift between 300 and 100~K ranges from approximately $2.16$ to $3.19~\mathrm{cm}^{-1}$, with negative temperature coefficients for all samples. The magnitude of $\chi$ is greatest for S1 and lower for the PVP-containing samples, although it does not vary monotonically with PVP content. Similarly, the TO-like band positions exhibit sample-dependent variations without a systematic shift across the PVP series. These observations indicate that PVP addition is accompanied by changes in the vibrational response of ZnS.

The persistence of the TO-like and LO-like features throughout the measured temperature range is consistent with the retention of the cubic ZnS structure identified by XRD. The characteristic sulfur-associated bands observed in S1 suggest a possible contribution from residual elemental sulfur, whereas their weaker or poorly resolved appearance in the PVP-containing samples indicates differences in their Raman response without confirming a reduction in sulfur content. Overall, the temperature-dependent Raman results demonstrate the preservation of the characteristic ZnS vibrational framework, accompanied by sample-dependent variations in phonon temperature coefficients and local vibrational environments, with no evidence of a temperature-induced structural transition between 100 and 300~K.

\subsection{Temperature-Dependent Photoluminescence Spectroscopy}
\label{subsec:temperature_pl}

The preceding Raman analysis showed that the characteristic ZnS lattice vibration is retained in S1--S4. Although the TO-like mode hardens slightly on cooling in every sample, its room-temperature position and temperature coefficient vary non-monotonically with PVP loading. The monotonic decrease in the Tauc optical gap from 3.69~eV for pristine ZnS (S1) to 3.65, 3.60, and 3.58~eV for S2, S3, and S4, respectively, therefore cannot be explained by uniform softening of the Zn--S lattice. Temperature-dependent photoluminescence (PL) spectroscopy was used to examine how PVP affects the localized electronic states involved in radiative recombination.

\begin{figure*}[t]
\centering
\includegraphics[width=1\textwidth]
{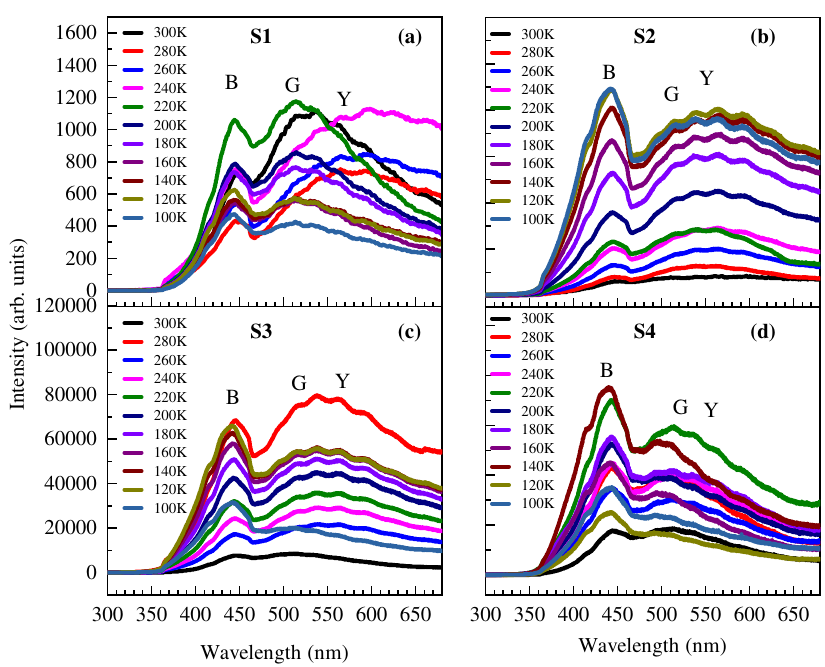}
\caption{Temperature-dependent PL spectra of (a) pristine ZnS (S1) and PVP-modified ZnS samples (b) S2, (c) S3, and (d) S4, recorded from 100 to 300~K in 20~K intervals.}
\label{fig:pl_temperature_spectra}
\end{figure*}

PL spectra were recorded from 100 to 300~K in 20~K steps and are presented in Fig.~\ref{fig:pl_temperature_spectra}. All four samples exhibit a broad, asymmetric visible-emission envelope rather than a narrow near-band-edge peak. A dominant blue maximum occurs near 440--449~nm, together with a broader lower-energy contribution extending through the green and yellow spectral regions.

The low-temperature spectra contain shoulders and secondary maxima within the broad visible-emission envelope. These features are consistent with partially resolved contributions from overlapping localized states, which become more apparent when thermal broadening and thermally activated non-radiative relaxation are reduced. Temperature-dependent studies of ZnS nanostructures have likewise shown different thermal responses for excitonic, surface-related, and defect-related emission bands \cite{Zhang2008ZnSNanocables,Ye2006,Wang2012}. In the present spectra, the comparatively stable blue maximum and sample-dependent lower-energy features indicate changes in the relative spectral weights of overlapping recombination channels rather than a set of independently shifting transitions. Small undulations that are not reproduced at neighbouring temperatures are therefore not assigned to additional defect levels.

The visible emission lies well below the absorption edge. The blue maximum at approximately 440--449~nm corresponds to an energy of about 2.77--2.82~eV, while the green--yellow contribution occurs at still lower energy. This separation between absorption and emission indicates that photoexcited carriers relax into localized sub-band-gap states before radiative recombination. Broad defect- and surface-related emissions have also been reported in ZnS nanostructures and capped ZnS nanoparticles \cite{Denzler1998ZnS,Ye2006,Wang2012,Ghosh2006PVPZnS}.

To separate the overlapping spectral contributions, each spectrum was fitted using four constrained Gaussian components together with a linear background. The short-wavelength shoulder, \(S\), was included to stabilize the fit near the edge of the analysis window and is not assigned a microscopic origin. The remaining components describe the dominant blue band (\(B\)), a shorter-wavelength part of the defect-emission envelope (\(G\)), and a longer-wavelength part (\(Y\)). Their fitted centre wavelengths were used to describe changes
in the positions of the overlapping spectral contributions.

Published studies provide more specific microscopic interpretations of these fitted spectral regions. Denzler \textit{et al.} resolved the dominant short-wavelength emission of colloidal ZnS nanocrystals into peaks at 416.1, 423.9, 430.1, and 437.8~nm and associated this fine structure with optical transitions involving zinc and sulphur vacancies and interstitials \cite{Denzler1998ZnS}. Within the commonly used defect-level picture, zinc vacancies (\(V_{\mathrm{Zn}}\)) and interstitial sulphur (\(I_{\mathrm{S}}\)) form acceptor-like states near the valence-band edge, whereas sulphur vacancies (\(V_{\mathrm{S}}\)) and interstitial zinc (\(I_{\mathrm{Zn}}\)) form donor-like states closer to the conduction-band edge \cite{Jindal2021ZnS}. Jindal and Sharma observed features near 359, 423, 451, 484, 528, and 567~nm under 260-nm excitation. They assigned the 359-nm feature to near-band-edge free-exciton recombination, the 423-nm emission to recombination from a sulphur-vacancy donor state to the valence band, and the 451- and 484-nm bands to recombination involving carriers trapped in shallow surface states \cite{Devarajan2012ZnS,Sooklal1996ZnS,Jindal2021ZnS}. The weaker features near 528 and 567~nm were associated with native defects, particularly zinc-vacancy-related acceptor states \cite{Dake2020CrZnS,Jindal2021ZnS}. On this basis, the present \(B\) component near 440--449~nm is compatible with overlapping vacancy-, interstitial-, and shallow-surface-trap emission, while \(G\) and \(Y\) describe progressively lower-energy parts of the defect-emission envelope that may include zinc-vacancy-related contributions. These comparisons provide physically plausible recombination pathways, but the substantial spectral overlap and absence of defect-specific measurements prevent a unique microscopic identification of the \(B\), \(G\), and \(Y\) components.

The \(B\), \(G\), and \(Y\) designations are therefore fitting labels rather than fixed physical bands associated with individual point defects. Vacancies, interstitials, surface states, and defect complexes can emit over overlapping wavelength intervals in ZnS \cite{Ye2006,Wang2012}. In particular, the \(G\) and \(Y\) components overlap strongly in several spectra.The fitted centres provide a descriptive summary of changes
in the overlapping emission envelope; they do not quantify
the relative spectral weights of its components or uniquely
identify individual defect transitions.

\begin{figure*}[t]
\centering
\includegraphics[width=0.97\textwidth]
{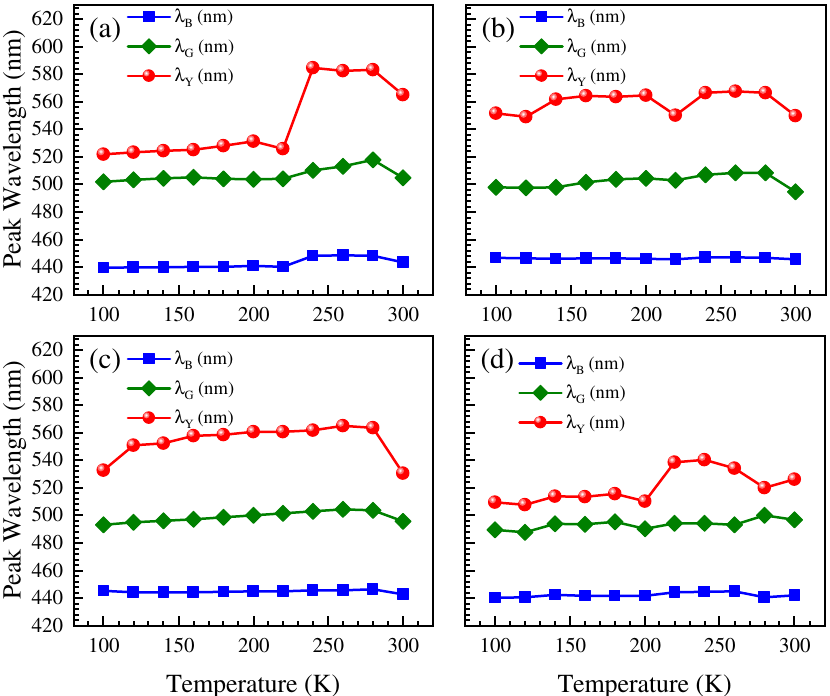}
\caption{Temperature dependence of the fitted centre wavelengths of the \(B\), \(G\), and \(Y\) emission components for (a) S1, (b) S2, (c) S3, and (d) S4. Lines are guides to the eye.}
\label{fig:pl_peak_wavelength_temperature}
\end{figure*}

The fitted centre wavelengths are presented in Fig.~\ref{fig:pl_peak_wavelength_temperature}. The \(B\)-component centre generally varies less than those of the lower-energy components, although S1 exhibits a distinct non-monotonic change above 220~K. The \(G\) and \(Y\) centres undergo larger apparent movements, but the strong overlap between these fitted components limits their interpretation as independently resolved emission bands.

For S1, the fitted \(B\)-component centre remains within 439.65--440.85~nm from 100 to 220~K. It increases to 448.16, 448.66, and 448.42~nm at 240, 260, and 280~K, respectively, before decreasing to 443.75~nm at 300~K. The \(G\) centre lies between 501.79 and 505.18~nm over 100--220~K, increases to 510.32--517.97~nm over 240--280~K, and decreases to 504.78~nm at 300~K. The fitted \(Y\) centre varies between 521.79 and 531.47~nm over 100--220~K, shifts to 582.49--584.70~nm over 240--280~K, and decreases to 565.22~nm at 300~K. The \(B\) component remains resolved throughout the temperature series, whereas \(G\) is weakly resolved from 240 to 300~K; \(Y\) is boundary-limited at 100--160~K and weakly resolved at 280~K. Accordingly, the large, non-monotonic changes in the fitted \(G\) and \(Y\) centres should be interpreted primarily as a redistribution of spectral weight within the overlapping lower-energy emission envelope rather than as continuous thermal shifts of independently resolved defect transitions.

For S2, the \(B\)-band centre remains within 445.57--447.37~nm throughout the temperature series. Up to 280~K, the \(G\) centre lies mainly between 497.53 and 508.59~nm, while the \(Y\) centre lies between 549.19 and 567.59~nm. Neither lower-energy component exhibits a consistent monotonic displacement. Thus, the dominant blue-emission energy is comparatively insensitive to temperature even though the shape of the measured emission envelope changes considerably on heating.

For S3, the \(B\) centre is 442.45~nm at 100~K and remains between 444.12 and 446.44~nm from 120 to 300~K. The fitted \(G\) centre changes from 493.17~nm at 100~K to 504.51~nm at 260~K, while \(Y\) lies mainly between 550.96 and 565.25~nm over 120--280~K. The \(G\) and \(Y\) contributions overlap substantially over parts of the temperature series. Changes in their fitted centres therefore primarily describe redistribution of spectral weight between overlapping emission regions rather than independent shifts of two resolved transitions.

For S4, the \(B\)-band centre remains within 440.34--444.95~nm, while the fitted \(G\) and \(Y\) centres span 487.92--500.19 and 507.92--540.39~nm, respectively. Here, \(Y\) denotes the longer-wavelength fitted component and should not be interpreted as a literal yellow-emission band over the entire temperature range. The strong overlap between \(G\) and \(Y\), together with the weak resolution of \(G\) at several temperatures, prevents their fitted movements from being interpreted as two independent thermal shifts.

Across S1--S4, the blue component remains within approximately 440--449~nm over most of the investigated temperature range and exhibits no common monotonic temperature dependence. In contrast, the lower-energy part of the spectrum changes substantially in shape and fitted decomposition. Temperature therefore affects the occupation and relative contributions of localized states more strongly than it changes the energy of the dominant blue-emission channel.

\begin{figure*}[htbp]
\centering
\includegraphics[width=1\textwidth]
{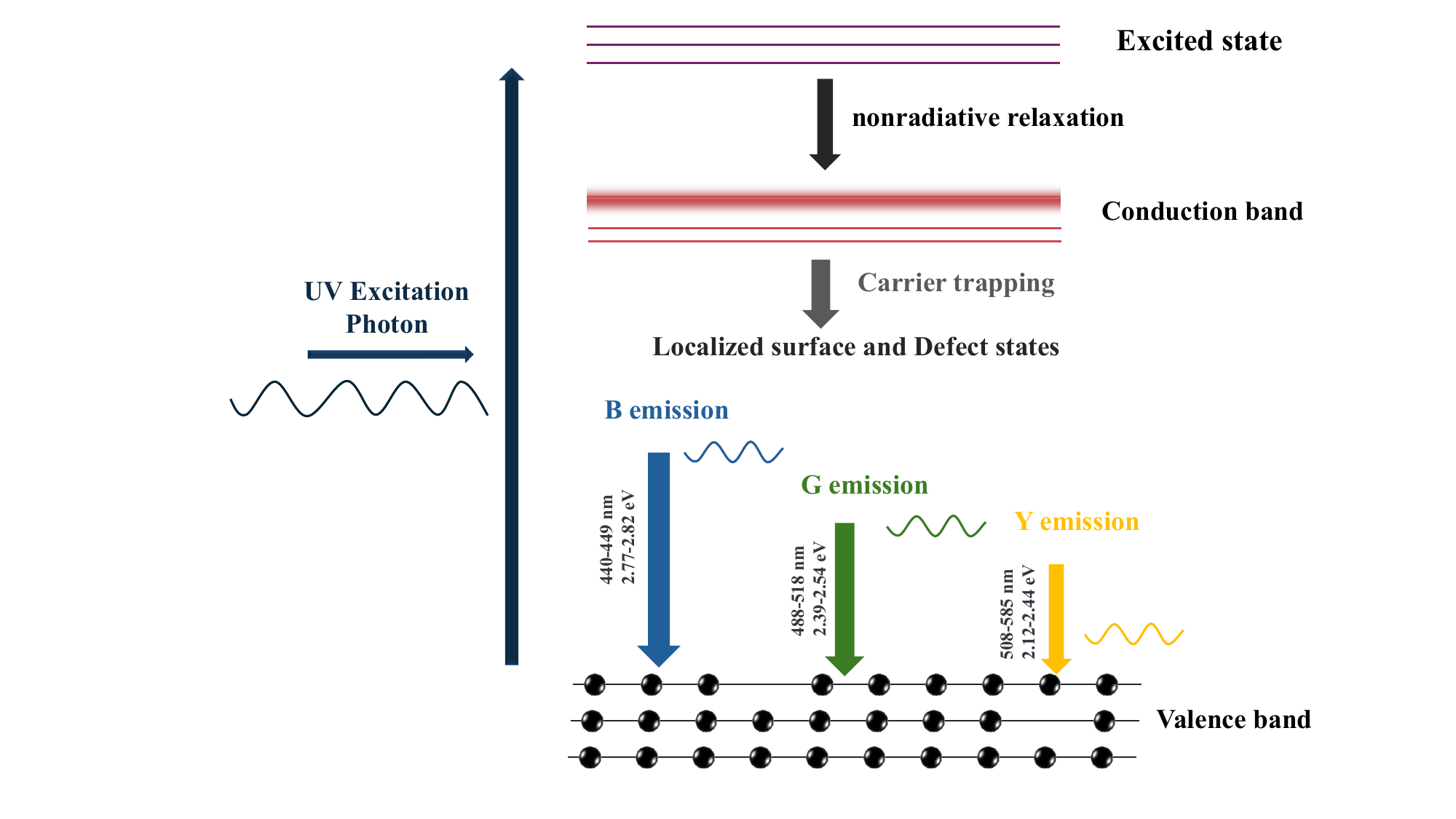}
\caption{Schematic illustration of carrier excitation, relaxation, trapping, and defect-mediated radiative recombination in pristine and PVP-modified ZnS nanoparticles. The \(B\), \(G\), and \(Y\) labels denote fitted regions within the broad PL envelope rather than three discrete defect levels.}
\label{fig:pl_pathway_schematic}
\end{figure*}

Figure~\ref{fig:pl_pathway_schematic} summarizes the carrier processes consistent with the PL measurements. Absorption of an ultraviolet photon generates an excited electron--hole pair, after which the carriers can undergo non-radiative relaxation, become trapped in localized states, or transfer between states before recombination. Radiative recombination through a distribution of defect- and surface-related states produces the broad visible-emission envelope \cite{Fang2011,Suyver2001,Ye2006,Prasad2019}. The \(B\), \(G\), and \(Y\) labels are consequently placed on the emitted spectrum rather than represented as three discrete electronic levels. Steady-state PL alone cannot determine whether a particular fitted contribution originates from a sulphur vacancy, zinc vacancy, interstitial, dangling bond, or defect complex.

PVP can influence these processes through its effects on nanoparticle growth and surface coordination. Interaction of the pyrrolidone group with surface Zn sites can modify aggregation, surface passivation, local strain, and the population of radiative and non-radiative traps \cite{Ghosh2006PVPZnS,Koczkur2015PVP,Shahi2017}. PVP-related excitation and energy-transfer effects have also been reported in capped ZnS systems \cite{Manzoor2004}. This interpretation is consistent with the structural and vibrational measurements, which indicate that PVP modifies the nanoparticle surface environment without causing a systematic transformation of the cubic ZnS lattice.

Overall, the combined results support a surface- and defect-mediated interpretation of the optical response. PVP modifies the distribution and radiative activity of localized states, while temperature changes their relative occupation and the competition between radiative and non-radiative relaxation. The principal PL result is therefore not a uniform thermal displacement of a single emission line, but a sample-dependent redistribution of emission across overlapping blue and lower-energy recombination channels.
\subsection{Photocatalytic degradation of methylene blue}
\label{sec:photocatalytic_degradation}

The preceding structural and optical results show that PVP-assisted preparation modifies the particle assembly and surface environment of ZnS while retaining its principal cubic phase. The photocatalytic response of the series was therefore evaluated using methylene blue (MB) as a model organic dye, with the visible absorption band of MB monitored during UV irradiation. Pristine ZnS (S1) was compared with ZnS prepared using 0.5, 1.5, and 2.5~g of PVP (S2, S3, and S4, respectively). Photocatalytic dye removal has previously been reported for PVP-capped ZnS, while comparative studies of differently capped ZnS nanocrystals have shown that surface modification can influence their photocatalytic response \cite{Deka2019ZnSCapping}. 

For each catalyst-containing measurement, 50~mg of powder was added to 100~mL of aqueous MB solution (5~mg\,L$^{-1}$), corresponding to a catalyst loading of 0.50~g\,L$^{-1}$. The suspensions were maintained in the dark for 30~min and subsequently exposed to 254~nm UV irradiation. Aliquots were collected at 30~min intervals up to 210~min and centrifuged to remove the catalyst particles, after which the supernatants were analyzed by UV--visible spectroscopy. An MB solution without catalyst was subjected to the same dark-contact and irradiation schedule as a catalyst-free control. The absorbance measured immediately after the dark-contact period was defined as $t=0$ for each irradiation series.

\begin{figure}[htbp]
\centering
\includegraphics[width=0.5\linewidth]{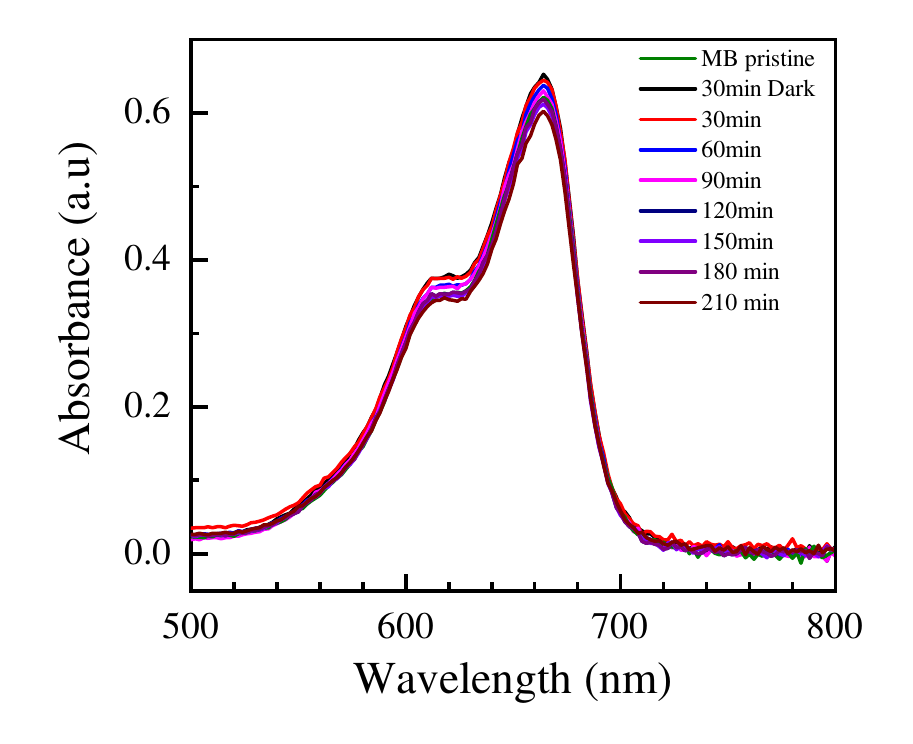}
\caption{UV–visible spectra of the catalyst-free MB solution before dark contact, after 30 min of dark contact, and during subsequent 254 nm UV irradiation (30–210 min).}
\label{fig:mb_pristine}
\end{figure}

The pristine MB solution in Fig.~\ref{fig:mb_pristine} exhibits a dominant absorption band near 664~nm together with a weaker shoulder near 610--620~nm. The 664~nm band was selected as the principal monitor of the MB response during irradiation, while the complete spectra were retained to examine changes in the band profile. This distinction is important because photooxidative N-demethylation of MB can produce hypsochromic shifts in its absorption spectrum during transformation \cite{Zhang2001MBDemethylation}. Consequently, a decrease in absorbance at a single wavelength cannot, by itself, identify the transformation products or establish mineralization. The pristine spectrum is distinct from the catalyst-free irradiation series, which measures MB-band attenuation in the absence of ZnS. The catalyst-free control exhibited 6.58\% apparent MB-band loss after 210~min. Karim \textit{\textit{et al}}.\ reported approximately 5\% MB removal after 180~min under 254~nm UV irradiation alone, while Zhao \textit{et al}.\ reported approximately 12\% in their blank MB control after 200~min under 254~nm UV irradiation \cite{Karim2022MBPhotolysis,Zhao2017MBControl}. 

\begin{figure}[htbp]
\centering
\includegraphics[width=0.97\linewidth]{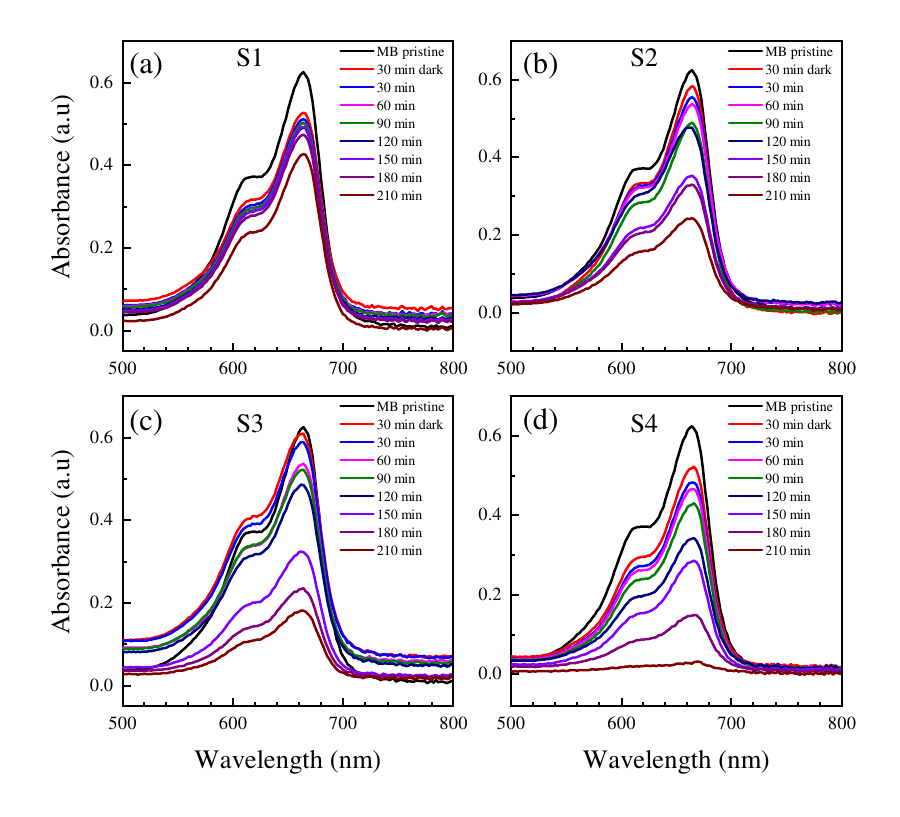}
\caption{Time-dependent UV--visible spectra of MB in the presence of (a) pristine ZnS (S1), (b) ZnS prepared with 0.5~g PVP (S2), (c) ZnS prepared with 1.5~g PVP (S3), and (d) ZnS prepared with 2.5~g PVP (S4).}
\label{fig:MB_spectra}
\end{figure}

Figure~\ref{fig:MB_spectra} presents the time-dependent UV--visible spectra obtained in the presence of S1--S4. The MB band near 664~nm remains detectable throughout the S1 series, whereas the PVP-modified samples exhibit progressively greater attenuation toward the later stages of irradiation. The pristine-MB trace included in each panel provides a spectral reference, while the post-dark spectrum of each catalyst suspension defines the starting point for its corresponding irradiation series.

Before dark contact, the MB solution used for the catalyst-containing suspensions had an absorbance of 0.6259 at 664~nm. After 30~min of dark contact, the absorbances were 0.5271, 0.5839, 0.6103, and 0.5201 for S1--S4, respectively. These differences indicate sample-dependent changes in the measured MB signal during dark contact, which may include dye uptake and other optical contributions. The pre-dark absorbance was therefore not used as a common reference for the irradiation experiments. Instead, $A_0$ denotes the post-dark absorbance of each individual suspension and $A_t$ denotes its absorbance after $t$ minutes of UV exposure. If the monitored absorption is dominated by the parent dye, the normalized absorbance provides an approximate measure of its relative concentration:

\begin{equation}
\frac{C_t}{C_0} \simeq \frac{A_t}{A_0}.
\label{eq:ctc0}
\end{equation}

The ratio $A_t/A_0$ is therefore used below as an absorbance-based measure of the MB-band response. For the catalyst-free control, the post-dark absorbance of 0.6439 was used as $A_0$, compared with a pre-dark absorbance of 0.6494. Thus, each irradiation series was normalized independently to its corresponding post-dark value.

\begin{figure}[htbp]
\centering
\includegraphics[width=\linewidth]{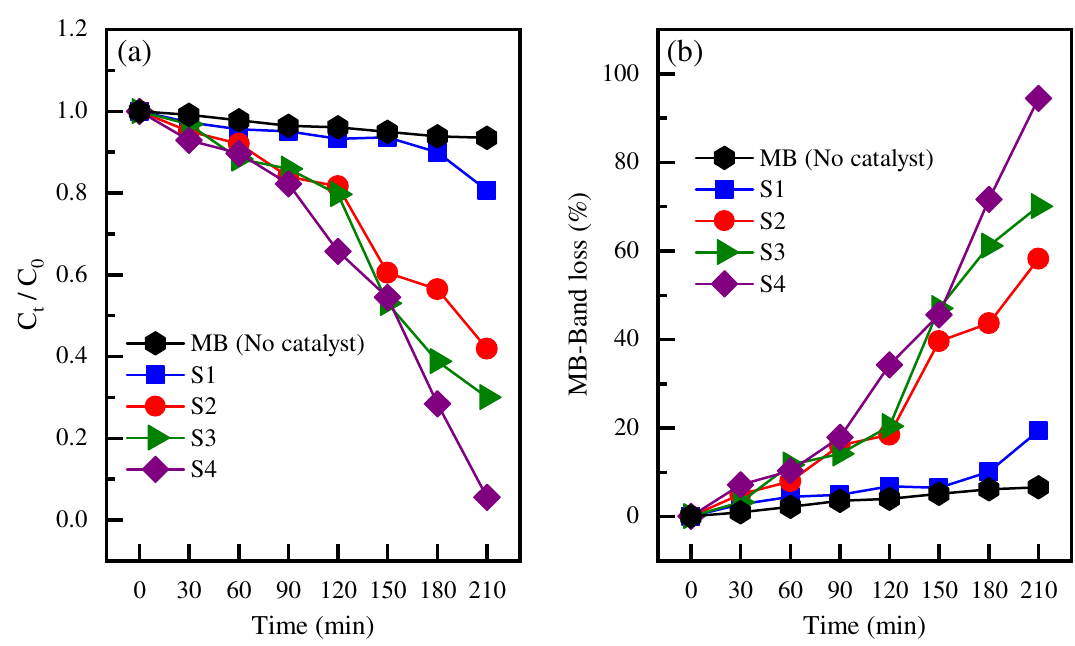}
\caption{Time-dependent UV-assisted methylene blue (MB)
band attenuation for pristine ZnS (S1), PVP-modified ZnS
nanoparticles (S2--S4), and the catalyst-free MB control
under 254~nm irradiation:
(a) normalized MB-band absorbance
($A_t/A_0 \simeq C_t/C_0$) as a function of irradiation
time; (b) corresponding apparent MB-band loss (\%),
calculated as $100(1-A_t/A_0)$.}
\label{fig:degradation}
\end{figure}

The normalized profiles are shown in Fig.~\ref{fig:degradation}(a).
At 210~min, the $A_t/A_0$ values are 0.8067, 0.4182, 0.3000,
and 0.0558 for S1--S4, respectively, compared with 0.9342 for
the catalyst-free control. All four ZnS-containing suspensions
therefore exhibit greater MB-band attenuation than the
catalyst-free control. 

The absorbance change can be expressed as the apparent
percentage loss of the principal MB band:

\begin{equation}
L_t(\%)=100\left(1-\frac{A_t}{A_0}\right).
\label{eq:loss}
\end{equation}

Figure~\ref{fig:degradation}(b) presents this response in
percentage form; it is calculated directly from the normalized
absorbance and is therefore not an independent measure of
degradation. At 150~min, S3 and S4 show similar apparent MB-band losses
of 47.06\% and 45.55\%, respectively. The S4 loss then
increases to 71.66\% at 180~min and 94.42\% at 210~min,
exceeding S3 at both times and giving the largest final
response of the series. The largest measured change in S4
occurs between 150 and 180~min. Because measurements were
acquired at 30~min intervals, the data do not establish
when within that interval the change occurred.

At the end of irradiation, the absorbances of S1--S4 are
0.4252, 0.2442, 0.1831, and 0.0290, respectively. Relative
to their individual post-dark absorbances, these values
correspond to apparent MB-band losses of 19.33\%, 58.18\%,
70.00\%, and 94.42\%. The MB-only control exhibits 6.58\%
apparent band loss. These values quantify attenuation of the
monitored absorption band and should therefore be regarded
as apparent MB-band losses rather than direct measurements
of the concentration of intact MB molecules. In particular,
they do not establish complete oxidation or mineralization
of MB
\cite{Houas2001MB,Chong2010WaterTreatment,Zhang2001MBDemethylation}.
In a different MB photooxidation system, Ward
\textit{et al}.\ used total organic carbon measurements to
distinguish decolorization from subsequent mineralization,
illustrating why loss of visible absorbance and carbon
removal should be assessed separately
\cite{Ward2013MBOxidation}.

A 254~nm photon carries approximately 4.88~eV, exceeding
the measured Tauc band gaps of all four samples:
3.69~eV for S1, 3.65~eV for S2, 3.60~eV for S3, and
3.58~eV for S4. Photoexcitation of ZnS is therefore
energetically accessible throughout the series. The modest
band-gap differences alone cannot account for the differences
in final MB-band loss or the distinct time-dependent
responses. Particle assembly, surface coordination, access
to reactive sites, and carrier trapping may also influence
the photocatalytic response
\cite{Ajibade2022ZnSMB,Jubeer2023Defect,Kaur2024ZnSWater,Soltani2012PVPZnS}.

Initial photoexcitation may be represented schematically as

\begin{equation}
\mathrm{ZnS}+h\nu
\longrightarrow
e^{-}_{\mathrm{CB}}+h^{+}_{\mathrm{VB}},
\label{eq:excitation}
\end{equation}

where $e^{-}_{\mathrm{CB}}$ and $h^{+}_{\mathrm{VB}}$
represent photogenerated conduction-band electrons and
valence-band holes. These carriers may reach the surface,
become trapped by localized states, or recombine. Their
contribution to interfacial reactions therefore depends
on the competition among these processes
\cite{Mills1997Semiconductor,Herrmann1999Heterogeneous,Schneider2014Photocatalysis}.

At the surface, conduction-band electrons may reduce
dissolved or adsorbed oxygen, whereas valence-band holes
may oxidize adsorbed hydroxide or water. Representative surface reactions can be represented as

\begin{equation}
\mathrm{O}_{2(\mathrm{ads})}+e^{-}_{\mathrm{CB}}
\longrightarrow
\mathrm{O}_{2}^{\bullet-},
\label{eq:superoxide}
\end{equation}

\begin{equation}
\mathrm{OH}^{-}_{(\mathrm{ads})}+h^{+}_{\mathrm{VB}}
\longrightarrow
{}^{\bullet}\mathrm{OH}_{(\mathrm{ads})},
\label{eq:hydroxyl1}
\end{equation}

and

\begin{equation}
\mathrm{H_2O}_{(\mathrm{ads})}+h^{+}_{\mathrm{VB}}
\longrightarrow
{}^{\bullet}\mathrm{OH}_{(\mathrm{ads})}+\mathrm{H}^{+}.
\label{eq:hydroxyl2}
\end{equation}

Photogenerated holes may also participate directly in dye oxidation.
These reactions illustrate possible photocatalytic pathways, but the
present measurements do not identify the reactive species involved
\cite{Houas2001MB,Ajibade2022ZnSMB,Jubeer2023Defect,
Mills1997Semiconductor}.

The sample-dependent response can also be considered in
terms of the PVP-modified interface. FTIR identifies
PVP-related groups in S2--S4, SEM reveals differences
in aggregate structure, and XRD and Raman show that
the principal cubic ZnS phase and its characteristic
vibrational response are retained. PVP-assisted
preparation may therefore change particle--solution
interactions and the accessibility of the ZnS surface.
The present measurements do not determine the surface
coverage of PVP or isolate the contribution of any
individual interfacial property.

The stronger MB-band attenuation in the PVP-modified
samples is consistent with earlier comparisons of
capped and uncapped ZnS. Kaur \textit{et al}.\ reported
faster MB removal with PVP-ZnS than with pristine ZnS
under UV irradiation, with the reported times being
60 and 120~min, respectively
\cite{Kaur2024ZnSWater}. Soltani \textit{et al}.\ also compared bare and
PVP-capped ZnS for MB removal and reported an improved
initial response after PVP coating under visible light
\cite{Soltani2014ZnSCoating}. Kumar \textit{et al}.\ reported
87\% removal with PVP-capped ZnS compared with 80\%
for uncapped ZnS under sunlight; their test dye was
methyl orange rather than MB \cite{Kumar2022PVPCapped}.
Deka \textit{et al}.\ also found that the MB response of
ZnS depends on the choice of capping agent
\cite{Deka2019ZnSCapping}. These studies support the
relevance of PVP and surface modification to dye
removal, although differences in dye, irradiation
conditions, and preparation method prevent direct
comparison of their percentages with the present
results.

The pronounced S4 change between 150 and 180~min
raises the possibility that conditions at the
particle--solution interface evolve during irradiation.
Svensson \textit{et al}.\ observed an induction period
during phenol photodegradation on ZnO and related it
to surface residues and the availability of active
sites \cite{Svensson2025ZnOInduction}. Horikoshi
\textit{et al}.\ demonstrated photocatalytic oxidation
of PVP at a TiO$_2$/water interface
\cite{Horikoshi2001PVPPhotomineralization}. Ward
\textit{et al}.\ reported irradiation-dependent
development of active sites in a bismuth-impregnated
silica system containing H$_2$O$_2$
\cite{Ward2013MBOxidation}. These examples show that
photocatalytic responses can change with exposure
time. Their materials and reaction conditions differ
from the present suspension, so they do not establish
an induction mechanism, PVP removal, or the cause of
the late S4 response.

Changes in dye adsorption, surface accessibility, or
the relative contribution of absorbing species could
also influence the measured 664~nm signal. In
particular, MB N-demethylation can change its
absorption profile during transformation
\cite{Zhang2001MBDemethylation}. The measured profiles
establish that S4 has the largest final apparent
MB-band loss, reaching 94.42\% after 210~min relative
to its post-dark starting value. They do not identify
the reactive species, determine why S4 changes most
strongly between 150 and 180~min, or demonstrate
mineralization
\cite{Houas2001MB,Zhang2001MBDemethylation,
Ward2013MBOxidation}.
\section{Conclusion}

Pristine ZnS (S1) and three samples prepared with increasing PVP addition
(S2--S4) were compared using structural, microscopic, vibrational, optical,
and UV-assisted MB measurements. XRD identified the cubic zinc-blende ZnS
phase throughout the series. The apparent Scherrer coherent-domain sizes
showed a non-monotonic variation with PVP addition, indicating that PVP
does not produce a uniform change in coherent-domain development. FTIR
confirmed the presence of PVP-related surface-associated groups, while SEM
revealed changes in aggregate morphology and surface texture with increasing
PVP addition.

TEM analysis of the end-member samples S1 and S4 revealed locally
crystalline regions within larger agglomerated structures. The TEM images
showed comparatively larger structural features for S1 than for S4,
consistent with the finer features observed for S4 in the SEM images. The
TEM-derived size distributions likewise indicated a smaller characteristic
crystallite dimension for S4 than for S1. The SAED ring assignments agreed
with the principal (111), (220), and (311) reflections of cubic
zinc-blende ZnS identified by XRD, confirming retention of the ZnS
crystalline framework. 

The allowed-direct-transition Tauc optical gap decreased from 3.69 to
3.58~eV between S1 and S4. The principal TO-like Raman band hardened on
cooling across the series, while the magnitude of its temperature response
varied between samples. Visible PL remained broad, with stronger
temperature- and sample-dependent redistribution in the overlapping
green--yellow region than in the blue region. These observations indicate
that PVP modification alters the optical and vibrational responses through
changes in the surface and localized-state environment while retaining the
principal cubic ZnS phase.

The UV-assisted MB measurements showed distinct time-dependent responses
among the four samples under 254~nm irradiation. After 210~min, the
apparent MB-band losses were 19.33\%, 58.18\%, 70.00\%, and 94.42\% for
S1--S4, respectively, compared with 6.58\% for the catalyst-free MB
control. The corresponding \(C_t/C_0\) values at 210~min were 0.8067,
0.4182, 0.3000, and 0.0558 for S1--S4, respectively. The time-dependent profiles showed
sample-dependent changes in the extent and rate of MB-band attenuation,
with S4 reaching the largest endpoint attenuation. Since the 254~nm
photon energy exceeds the optical gaps of all four samples, photoexcitation
of ZnS is energetically accessible. However, the relatively small variation
in optical gap alone cannot account for the distinct time-dependent
responses. Differences in particle assembly, surface coordination,
accessible surface sites, and localized electronic states may contribute
to the observed sample-dependent behavior. The measured decrease in the
MB absorption band demonstrates chromophore attenuation under the
experimental conditions, but does not by itself establish complete
mineralization or identify the reaction products.

Overall, the combined structural, microscopic, vibrational, optical, and
UV-assisted measurements show that PVP modification preserves the cubic
zinc-blende ZnS crystalline framework while altering the surface
environment, aggregate morphology, local crystallite dimensions, optical
gap, temperature-dependent vibrational response, visible emission profile,
and UV-assisted MB-band response. The results therefore support a
PVP-dependent modification of the surface and local electronic environment
of ZnS rather than formation of a new crystalline phase.

\section*{Acknowledgments}

VSR and KRK gratefully acknowledge the Science and Engineering Research 
Board (SERB), Department of Science and Technology (DST), Government of 
India, for financial support under grant number EEQ/2022/000354. The authors 
also acknowledge the Central Instrumentation Facility at IIT Gandhinagar 
for providing access to instrumentation for various characterizations. 

\bibliography{references,sn-bibliography}

\end{document}